\documentclass[useAMS,usenatbib]{mn2e}

\usepackage{graphicx}
\usepackage{rotating}
\usepackage{longtable}
\usepackage{lscape}

\def\Msun{M$_\odot$ }
\def\Mbh{${\cal M}_{\rm BH}$}

\def\Mhost{${\cal M}_{\rm host}$}
\def\Edd{$L/L_{\rm Edd}$}
\def\lLl{$\lambda L_{\lambda}$}

\def\Civ{C\,{\sc iv}}
\def\Mgii{Mg\,{\sc ii}}
\def\FeII{Fe\,{\sc ii}}
\def\FeIII{Fe\,{\sc iii}}

\def\Hb{H$\beta$}
\def\lsim{\mathrel{\rlap{\lower 3pt \hbox{$\sim$}} \raise 2.0pt \hbox{$<$}}}
\def\gsim{\mathrel{\rlap{\lower 3pt \hbox{$\sim$}} \raise 2.0pt \hbox{$>$}}}

\title[The \Mbh--\Mhost{} relation through Cosmic Time]{
 The quasar \Mbh--\Mhost{} relation through Cosmic Time\\
 I -- Dataset and black hole masses
}
\author[Decarli et al.]{
R. Decarli$^{1,2}$\thanks{E-mail: decarli@mpia-hd.mpg.de},
R. Falomo$^{3}$,
A. Treves$^{1,4}$,
J.K. Kotilainen$^{5}$,
M. Labita$^{1}$,
and R. Scarpa$^{6}$
\\
$^{1}$ Universit\`{a} degli Studi dell'Insubria, via Valleggio 11,
             I-22100 Como, Italy\\
$^{2}$ Max Planck Institute f\"ur Astronomie, K\"onigsthul 17, D-69117 Heidelberg, Germany\\
$^{3}$ INAF - Osservatorio Astronomico di Padova, Vicolo dell'Osservatorio 5,
             I-35122, Padova, Italy\\
$^{4}$ Istituto Naz. Fis. Nucleare, Piazza della Scienza 3, I-20126 Milano, Italy\\
$^{5}$ Tuorla Observatory, Department of Physics and Astronomy,
University of Turku, V\"ais\"al\"antie 20, FI-21500 Piikki\"o, Finland\\
$^{6}$ Instituto de Astrofisica de Canarias, Via Lactea, s/n E-38205
La Laguna (Tenerife), Spain
}
\begin{document}

\date{ }

\pagerange{\pageref{firstpage}--\pageref{lastpage}} \pubyear{2009}

\maketitle

\label{firstpage}

\begin{abstract}
We study the \Mbh{}--\Mhost{} relation as a function of
Cosmic Time in a sample of 96 quasars from $z=3$ to the present
epoch. In this paper we describe the sample, the data sources and
the new spectroscopic observations. We then illustrate how we derive
\Mbh{} from single-epoch spectra, pointing out the uncertainties in
the procedure. In a companion paper, we address the dependence of
the ratio between the black hole mass and the host galaxy luminosity
and mass on Cosmic Time.
\end{abstract}


\section{Introduction}

The discovery of tight relations between the mass of massive black
holes, \Mbh{}, and the large scale properties of the galaxies where
they reside \citep[see][for a review]{ferrarese06} is one of the
most intriguing results in astrophysics of the last decade, given
the consequences in the frame of galaxy formation and evolution.
When and how these relations are set are still open questions.

Measuring \Mbh{} of quiescent massive black holes is extremely
challenging even in nearby galaxies, and has been done successfully
only in few tens of cases \citep{ferrarese06,pastorini07}. By
contrast, an estimate of \Mbh{} in Type-I Active Galactic Nuclei
(AGN) is possible by assuming that the gas emitting broad lines is
in virial equilibrium \citep[][but see also Marconi et al. 2008 and
references therein for possible effects of non-virial
components]{peterson00}, and that the line width traces the black
hole potential well. Based on reverberation mapping studies
\citep{blandford82}, \citet{kaspi00} found a correlation between the
characteristic size of the broad line region (hereafter, BLR) and
the continuum luminosity of the AGN. This allows an estimate of the
black hole mass from single epoch low-resolution spectra.

In order to sample the black hole -- host galaxy relations through
a wide range of Cosmic Ages, one has to focus on the brightest AGN.
Quasars have been detected up to $z\gsim6$
\citep[][]{fan04}. Large field surveys such as the SDSS allowed
a detailed spectroscopic study of $\sim60,000$ quasars with $z<4$ \citep{shen08}.
The drawback is that the typical nuclear-to-host luminosity ratio
in quasars is such that the light from the host galaxy is outshone
by the nuclear component. This usually prevents the detection of stellar
features in the spectra of bright quasars. Only through the excellent
resolution of the HST, together with state-of-art observing
techniques in the NIR at ground-based telescopes, the detection of
the extended emission of the host galaxies of few hundreds of quasars
up to $z\lsim3$ became possible \citep[see][and references therein]{kotilainen09}.

In this project we focus on quasars in the redshift range $0<z<3$ with
known host galaxy luminosity, in order to study the evolution of the
\Mbh{}--\Mhost{} relation. This will shed some light on the joint
evolution of black holes and galaxies up to and
immediately beyond the crucial age of maximum quasar activity
\citep{dunlop90} and star formation \citep{madau98}.
Here we present the sample, the new spectroscopic observations and
we infer the black hole masses.
In a companion paper \citep{paperII}, we address the topic of the
\Mbh--\Mhost{} relation as a function of Cosmic Time.


Throughout the paper, we adopt a concordance cosmology
with $H_0=70$ km/s/Mpc, $\Omega_m=0.3$, $\Omega_\Lambda=0.7$.
We converted the results of other authors to this cosmology when
adopting their relations and data.

\section[]{The sample}\label{sec_sample}

\begin{figure}
\begin{center}
\includegraphics[width=0.49\textwidth]{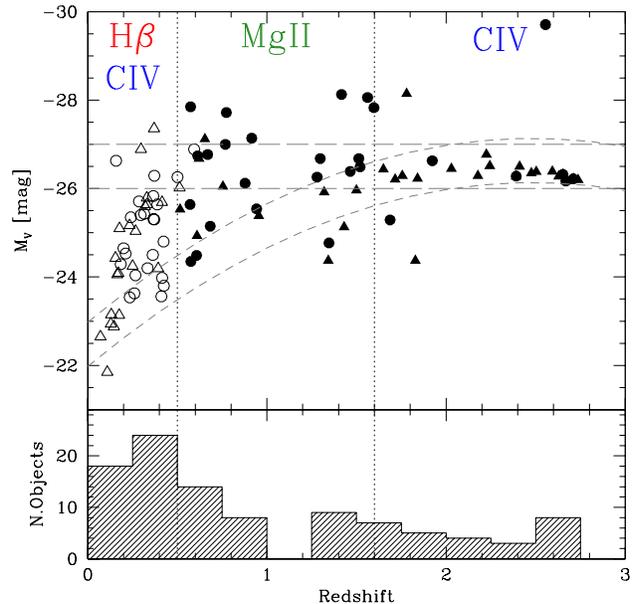}\\
\caption{The distribution of the quasars in our sample in the
($z$,$M_V$) plane. $M_V$ is the total rest-frame $V$-band absolute
magnitude of the quasars, estimated from the $V$-band apparent magnitudes
available in the \citet{vcv06} catalogue and $k$-corrected assuming the
quasar template by \citet{francis91}. Filled symbols refer to the mid- and
high-$z$ data for which we present optical spectroscopy in this paper. Empty
symbols mark the $z\lsim0.5$ data from \citet{labita06} and \citet{decarli08a}.
Circles are radio loud quasars, triangles are radio quiet. The broad
emission lines observed in the various redshift windows are also labelled.
We note that the $-27<M_V<-26$ luminosity range (long-dashed lines),
as well as the $M_*(z)>M_V>M_*(z)-1$ \citep[where $M_*(z)$ is the
characteristic luminosity of the quasar luminosity function by][plotted in
short-dashed lines]{boyle00} are well sampled at any redshift bin.
}\label{fig_distr_z}
\end{center}
\end{figure}

Up to now, few hundreds of quasar host galaxies have been resolved at $z<3$.
In order to minimize the uncertainties due to colour and filter corrections,
we select objects observed with filters roughly sampling the rest-frame
$R$-band. Furthermore, as the \Mbh{}--\Mhost{} relation is sensitive to the
spheroidal rather than the total mass of the host galaxy
\citep{marconi03,haring04}, we consider only those targets with host galaxies
classified as elliptical. Our sample consists of:
\begin{itemize}
\item[i--] 43 low redshift ($z\lsim0.5$) quasars imaged with the HST-Wide
Field Camera \citep{bahcall97,hooper97,boyce98,kirhakos99,hamilton02,pagani03,
dunlop03,floyd04,labita06,kim08a}. They represent the bulk of our
knowledge of low-$z$ quasar host galaxies. UV spectra of 28 of these quasars
are taken from the HST Faint Object Spectrograph archive \citep[see][]{labita06}.
For 36 objects we collect optical spectra from the Sloan Digital Sky Survey
\citep[][13 quasars]{adelman07} and through on-purpose observations taken at the
Asiago Telescope \citep[23 quasars; see][]{decarli08a}.
\item[ii--] 60 mid- and high-redshift ($z>0.5$) quasars imaged in the NIR
through ground-based observations under optimal seeing conditions
performed by our group \citep[50 objects; see][]{kotilainen98,kotilainen00,
falomo04,falomo05,hyvonen07a,hyvonen07b,kotilainen07,falomo08,kotilainen09,
decarli09a} or from
HST-based compilations \citep[10 sources from][]{kukula01,ridgway01}.
We note that all these studies have very high host galaxy detection rates
($>$$85$ per cent at $z<2$, $>$$60$ per cent beyond $z=2$). Optical
spectra were collected at the Nordic Optical Telescope and the ESO/3.6m
telescope (see section \ref{sec_observations}).
\end{itemize}
The 5 unresolved quasars in \citet{kotilainen09} were spectroscopically 
observed when the analysis of the imaging data was not complete yet. They 
are not included in the study of the evolution of the \Mbh{}--\Mhost{} 
relation.
Other two objects were dropped, as we could not estimate \Mbh{} from our
spectra (see Appendix \ref{sec_appendix}). Therefore the final sample consists 
of 96 quasars.
According to the \citet{vcv06} catalogue, 48 of them are radio loud quasars
(RLQs) and 48 are radio quiet (RQQs).
We remark that our sample is approximately twice as large as those of
\citet{peng06a,peng06b} and \citet{mclure06} and represents the largest
dataset ever considered in the study of the evolution of the \Mbh{}--\Mhost{}
relation. The distribution of our targets in the ($z$,$M_V$) plane is
shown in Figure \ref{fig_distr_z}. Table \ref{tab_sample} lists the main properties of each quasar in
our sample.

\section[]{New observations, data reduction}\label{sec_observations}

Spectra of the $z>0.5$ quasars were collected in 5 observing runs at the
European Southern Observatory (ESO) $3.6$m telescope in La Silla (Chile)
and at the $2.56$m Nordic Optical Telescope (NOT) in La Palma (Spain). Table
\ref{tab_runs} lists the dates of the observations and the number of spectra
collected in each run, while table \ref{tab_journal} summarizes the journal
of observations.

The ESO Faint Object Spectrograph and Camera (v.2, EFOSC2; see Buzzoni
et al. 1984) and its twin NOT instrument, the Andalucia Faint Object
Spectrograph and Camera (ALFOSC), were mounted in long-slit spectroscopy
configuration. EFOSC2 observations were carried out with grism \#4,
yielding a spectral resolution of $R\sim 400$ ($1.2"$ slit) in the
spectral range $4100$--$7500$ \AA{} ($\Delta\lambda/$pxl=$3.36$
\AA/pxl). For NOT observations, we used ALFOSC grisms \#6 and \#7,
which allow the observation of the $3500$--$5530$\footnote{The
nominal observed range of ALFOSC grism \#6 is larger blue-wards, but
the sensitivity is so low that we decided to drop the observed
spectra at wavelengths below $3500$ \AA.} and $3800$--$6840$ \AA{}
windows with spectral resolutions $R\sim490$ and $650$ with the
$1.0"$ slit ($\Delta\lambda/$pxl$\approx1.5$ \AA/pxl). At the central
wavelength, the instrumental resolutions are $12.6$ \AA{} (EFOSC2+grism
\#4) and $8.2$ \AA{} (ALFOSC+grism \#6 and \#7).
\begin{table}
\begin{center}
\caption{List of the observing runs. (1) Run ID. (2) Telescope
(3) Dates of observations. (4) Number of observed objects per run.
} \label{tab_runs}
\begin{tabular}{cccc}
   \hline
   Run & Telescope  & Nights            & N.obj.\\
   (1) &  (2)       &  (3)              & (4)   \\
   \hline
   E77 & ESO/$3.6$m & Sep.30-Oct.1 2005 &  6 \\
   E78 & ESO/$3.6$m & Mar.23-25    2007 & 12 \\
   E79 & ESO/$3.6$m & Sep.8-12     2007 & 22 \\
   N35 & NOT        & Apr.9-10     2007 &  2 \\
   N36 & NOT        & Oct.17-19    2007 & 18 \\
   \hline
   \end{tabular}
   \end{center}
\end{table}

\begin{figure*}
\begin{center}
\includegraphics[angle=-90, width=0.9\textwidth]{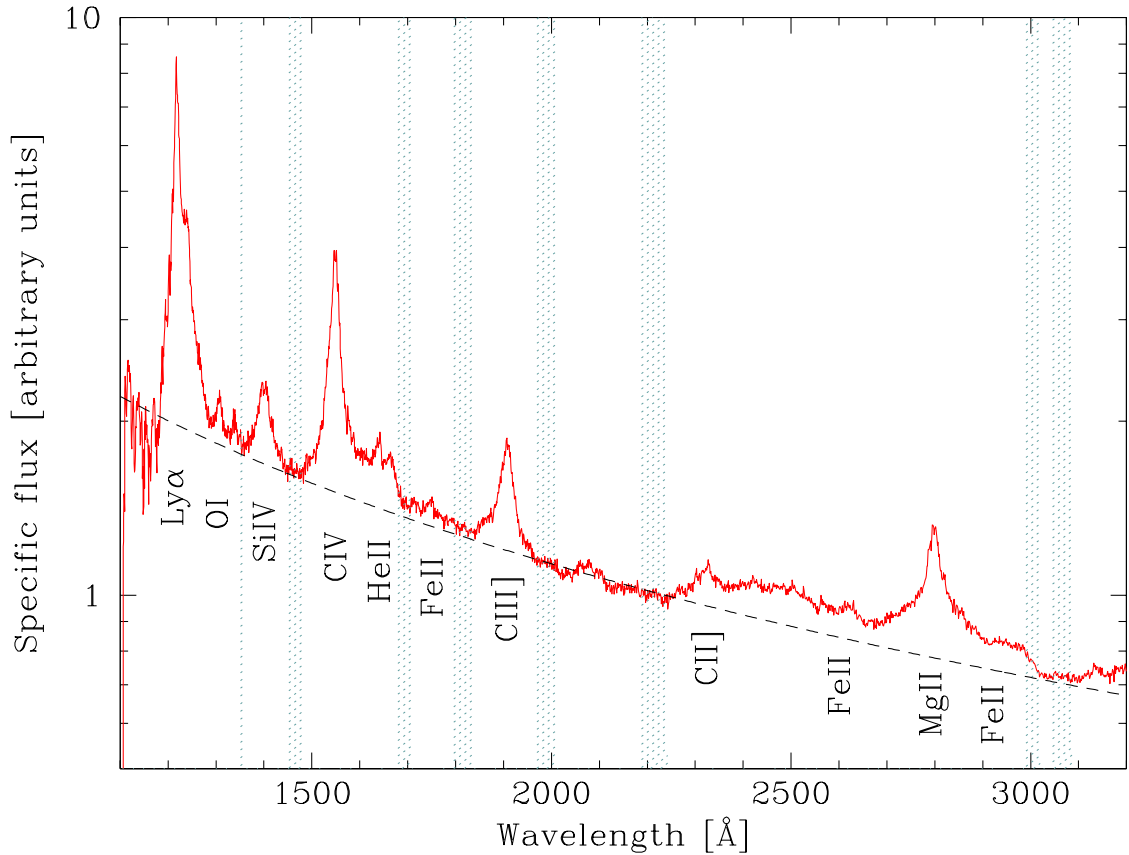}\\
\caption{The combined spectrum of the quasars considered in the
present study, normalized to the continuum value at 2250 \AA{}.
The median spectrum is plotted with a solid line. Main emission lines are
labelled. The shaded regions mark the intervals
used in the fit of the continuum. The resulting fit is plotted as a dash line.
}\label{fig_cont}
\end{center}
\end{figure*}

Standard IRAF tools were used in the data reduction. The
\texttt{ccdred} package was employed to perform bias subtraction, flat
field correction, image alignment and combination. Cosmic rays were
eliminated by combining 3 or more exposures of the same objects, and
applying \texttt{crreject} algorithm while averaging.
When only one or two bidimensional spectra were available, we applied
the \texttt{cosmicrays} task in the \texttt{crutils} package. In these
cases, in order to
prevent the task from altering the narrow component of emission
lines, we masked the central region of our bidimensional spectra.
The spectra extraction, the background subtraction and the
calibrations both in wavelength and in flux were performed with
\texttt{doslit} task in \texttt{specred} package, using He-Ar, Th-Ar and
He-Ne lamps and spectrophotometric standard stars as reference. Wavelength
calibration residuals are around $0.5$, $0.35$ and $0.03$ \AA{} in the
three adopted setups (sub-pixel), thus implying a negligible ($<1$ per
cent) error on redshift estimates.
Absolute flux calibration of spectra was corrected through the photometry
of field stars, as described in \citet{decarli08a}. This procedure
yields uncertainties in the flux calibration as low as $0.05$ mag
\citep[see, e.g., ][]{kotilainen09}, and commonly around $0.1$ mag.
Photometric accuracy of each target is reported in table \ref{tab_journal}.
Galactic extinction was accounted for according to \citet{schlegel98},
assuming $R_V = 3.1$.
We shifted the spectra to the rest frame, according to the catalogue
$z$, remembering that quasar lines with different ionization potentials
may present slightly different shifts \citep[e.g.][]{bonning07}.
Average signal to noise ratio of our spectra is $\sim$30.
The composite spectrum, obtained by median averaging rest frame individual
observations of these new data, is presented in Figure \ref{fig_cont}.
The whole dataset is available electronically at
\texttt{www.dfm.uninsubria.it/astro/caqos/}.

\section[]{Data analysis}\label{sec_analysis}

We focus our attention on the analysis required to estimate \Mbh{} from
single epoch observations of the rest-frame UV spectra of quasars. Applied
to the gas in the BLR of Type-1 AGN, the virial paradigm yields:
\begin{equation}\label{eq_virial}
{\cal M}_{\rm BH} = G^{-1} R_{\rm BLR} v_{\rm BLR}^2
\end{equation}
where $R_{\rm BLR}$ is the characteristic radius of the
broad line emission, and $v_{\rm BLR}$ is the velocity of
the emitting clouds at $R_{\rm BLR}$.

The cloud velocity can be inferred from the width of the broad emission
lines, e.g.:
\begin{equation}\label{eq_def_f}
v_{\rm BLR} = f \cdot {\rm FWHM}
\end{equation}
where $f$ is a geometrical factor around unity which accounts
for the de-projection of $v_{\rm BLR}$ from the
line-of-sight, and FWHM is the Full Width at Half Maximum of
the line profile (see McGill et al. 2008 and Decarli et al. 2008 for
discussions on other line width parametrizations). On the other hand,
the BLR size cannot be directly measured from single epoch
spectra. Our estimates of the broad line region size rely on the
discovery that $R_{\rm BLR}$ scales with a certain
power of the continuum luminosity of the AGN, $\lambda L_\lambda$
\citep[see][]{kaspi00}, as expected from simple photoionization models.

\subsection[]{The continuum luminosity and the BLR size}

Quasar UV--optical spectra are characterized by the superposition
of the following components:
\begin{itemize}
\item[-] A power-law-like continuum from the nucleus;
\item[-] Broad lines emitted within the BH influence radius;
\item[-] Narrow emission lines from the quasar host galaxy and the
AGN Narrow Line Region;
\item[-] The star light continuum from the host galaxy;
\item[-] A pseudo-continuum due to the blending of several \FeII{}
and \FeIII{} multiplets.
\end{itemize}
In order to infer the BLR radius, we have to isolate the first
component from the others. Our spectra cover the rest-frame UV
range of bright AGN, where the flux from the host galaxy star
light is always negligible.
The contamination of both broad and narrow emission lines is usually
avoided simply by fitting the power-law continuum to the
observed spectra in a number of wavelength windows free of strong
features. Here we adopted the intervals:
1351--1362, 1452--1480, 1680--1710, 1796--1834, 1970--2010, 2188--2243,
2950--2990, 3046--3090 \AA{} (see Figure \ref{fig_cont}). The fitted
parameters, together with the derived monochromatic specific fluxes and
luminosities are reported in table \ref{tab_fits}.
We note that the luminosity estimates obtained through the fit of
the power-law component and those obtained from a direct measure of
the continuum at 1350 and 3000 \AA{} are practically equivalent in
our datasets, the differences being $<10$ per cent on average.

Concerning luminosity--radius relations, simple photoionization
models with a constant ionization parameter predict that a ionizing
source emitting isotropically affects a region with a characteristic
radius scaling as the square root of the source luminosity. This
dependence has been confirmed in several reverberation mapping
studies focused on \Hb{}. Time-lag data of the rest-frame UV lines
are still limited in number, therefore any available relation for
\Mgii{} and \Civ{} is affected by poor statistics \citep[e.g., see
figure 6 in][]{kaspi07}. For consistency with our low-redshift
studies, we will refer to the the relations provided by
\citet{mclure02} for \Mgii{}\footnote{\citet{mclure02} set their
relation by comparing the \Hb{} time lags to the continuum
luminosity around 3000 \AA{}, given the fact that \Mgii{} and \Hb{}
have similar ionizing potentials, hence they are emitted in the same
region. This assumption allows us to adopt the same geometrical
factor for the \Mgii{} BLR as the one derived in \citet{decarli08a}
for \Hb{}. Note that \citet{mclure02} fitted the line profile of 
\Mgii{} into a broad and a narrow component, and used only the former to
infer \Mbh{} \citep[although there is no evidence of significant narrow
components in \Mgii{} lines: see, e.g.,][]{shen08}. As a consequence,
the values of \Mbh{} reported in their study are systematically larger 
than ours, but our conclusions are unchanged as we adopt our own internal
calibration of $f$.}:
\begin{equation}
\frac{R_{\rm BLR}({\rm MgII})}{10~ {\rm lt-days}}=(2.52\pm0.3) \left[\frac{\lambda L_{\lambda}(3000 \AA)}{10^{44} {\rm erg/s}}\right]^{0.47\pm0.05}
\end{equation}
and by \citet{kaspi07} for \Civ{}:
\begin{equation}
\frac{R_{\rm BLR}({\rm CIV})}{10~ {\rm lt-days}}=(0.24\pm0.06) \left[\frac{\lambda L_{\lambda}(1350 \AA)}{10^{43} {\rm erg/s}}\right]^{0.55\pm0.04}
\end{equation}
All these relations are based on low-redshift ($z<0.3$) objects,
with the only exception of S5 0836+71 \citep[$z=2.17$; see][]{kaspi07}.
Throughout this work, we assume that no significant Cosmic evolution of
the radius--luminosity relations occurs in the sampled redshift range.

The \Civ{} relation is poorly constrained, as mentioned above,
especially in the bright side where most of our objects reside.
Assuming a different slope of the luminosity--time lag relation,
e.g., $0.5$, affects our radius estimates at \lLl $\sim$ $10^{47}$
erg/s up to $0.2$ dex (i.e., a factor $\approx1.7$). Nevertheless,
we stress here that in terms of \Mbh{} estimates such a difference
would result in a global offset of all values and in a re-definition
of the geometrical factor $f$, with little effect on the evolution
of the \Mbh{}--\Mhost{} relation.

The scatter around the luminosity--radius relations dominates
the uncertainties in the radius estimates, contributing up to a
factor $\sim2$, while the uncertainties in the luminosity
estimates never exceed 10 per cent.


\begin{figure}
\begin{center}
\includegraphics[width=0.49\textwidth]{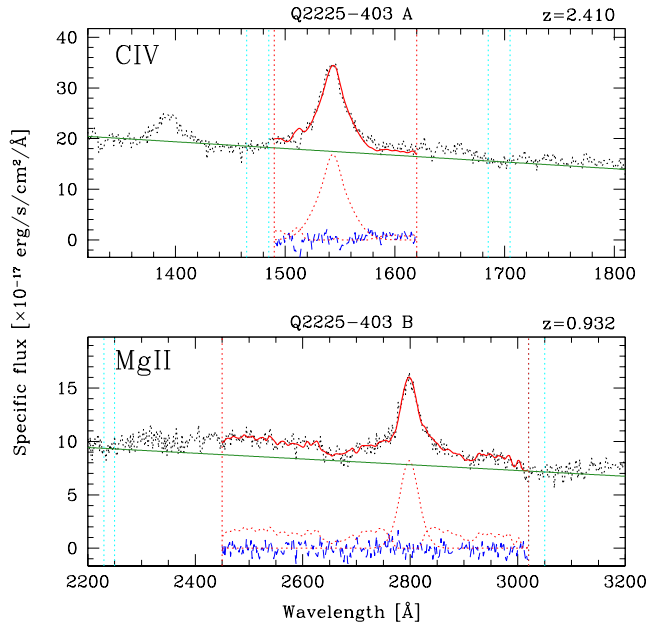}\\
\caption{Examples of the line fit, for \Civ{} in quasar
Q2225-403 A (\emph{upper panel}) and for \Mgii{} in quasar
Q2225-403 B (\emph{lower panel}). The rest-frame observed
spectra of the two quasars are shown in dotted lines. Dotted
vertical lines mark the regions used for the
underlying continuum estimates and the fitted wavelength range. The
thin, solid lines show the continua underlying broad emission lines.
The bold, solid lines are the broad emission line fits. In the bottom
side of each panel, we plot the fitted \FeII{} template and the broad
line model (dotted lines) together with the fit residual (dashed line).
}\label{fig_fit}
\end{center}
\end{figure}

\subsection[]{Emission line widths and cloud velocities}

Measuring the width of broad lines consists of three key steps: the
definition of the local continuum, the removal of contaminating spectral
features (in particular the \FeII{} multiplets), and the fit with a
certain analytical function.

We define underlying continua by matching the fluxes in the windows
2230--2250 and 3020--3050 \AA{} for \Mgii{} and 1465--1485 and 1685--1705 \AA{}
for \Civ{}, where no significant feature is observed (see Figure \ref{fig_fit}).
To evaluate the effects of \FeII{} contamination, we first fitted the \Mgii{}
and \Civ{} lines in the 2720--2880 and 1490--1570 \AA{} windows, where the
\FeII{} contribution is less relevant. Then, we compared these estimates of the
line widths with those obtained by fitting the 2450--3020 and 1490--1620 \AA{}
regions with a superposition of a broad line model plus a template reproducing
the \FeII{} emission \citep{vestergaard01}. In the latter approach the \FeII{}
emission observed in narrow line Type-1 AGN (usually, IZw001) is taken as
representative for all quasars, after being properly broadened. The validity
of this assumption has been widely discussed in the literature, both from a
theoretical and an observational point of view
\citep[e.g.,][]{phillips78,boroson92,marziani03,tsuzuki06,mcgill08}.
Since most \FeII{} lines are blended with each others, their width
is hardly constrained. We therefore fixed their broadening to the width of
\Civ{} and \Mgii{} lines. Figure \ref{fig_fit} shows examples of fits on
\Civ{} and \Mgii{}. Line width estimates obtained with or without the
\FeII{} modelling are usually consistent within 20 per cent, but
\Mgii{} shows few deviations larger than 30 per cent, especially when the
\FeII{} emission is strong with respect to the line flux. Hereafter, we
will refer to the line models obtained by template fitting the \FeII{}
emission.

\begin{figure*}
\begin{center}
\includegraphics[width=0.4\textwidth]{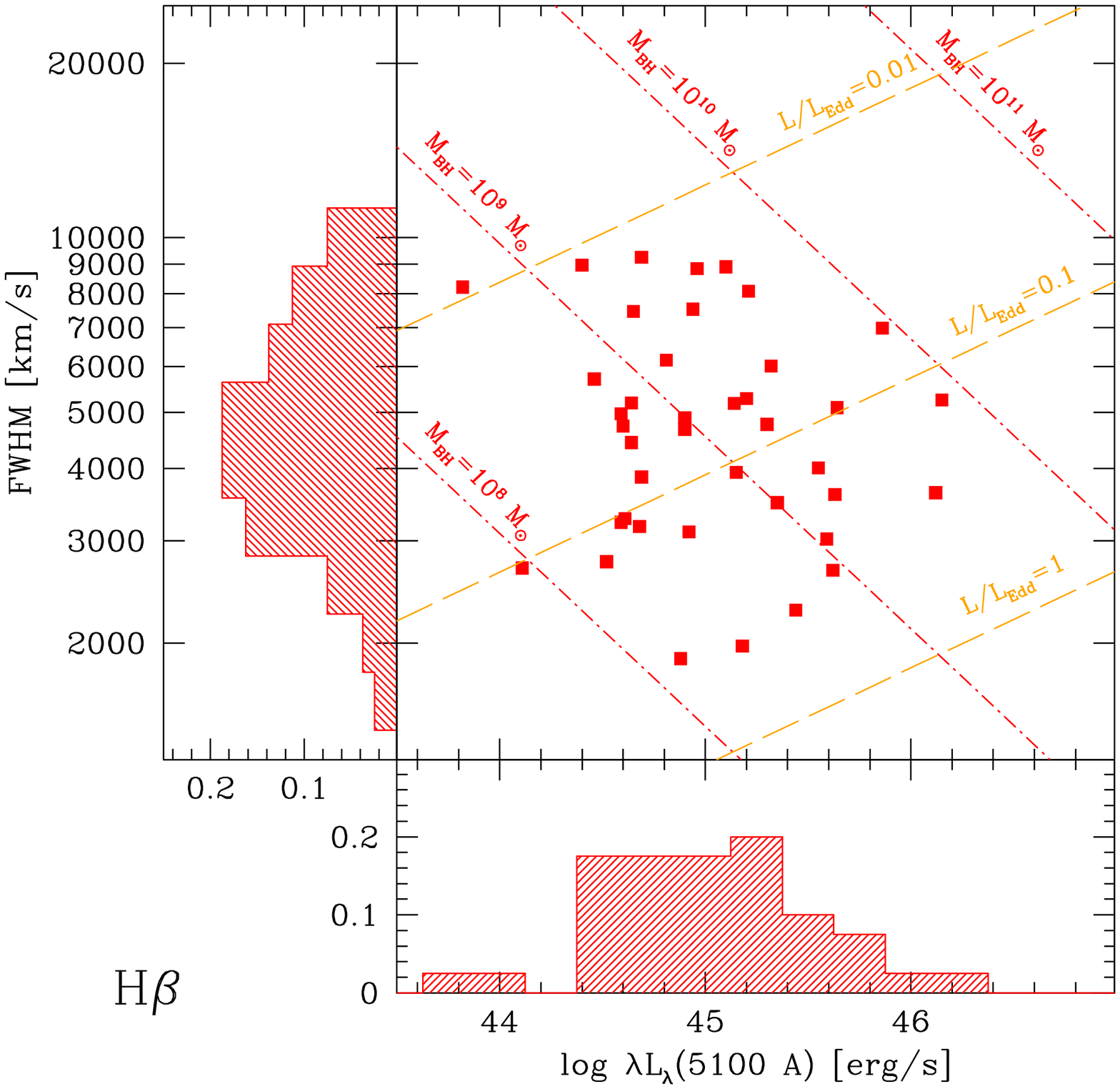}
\includegraphics[width=0.4\textwidth]{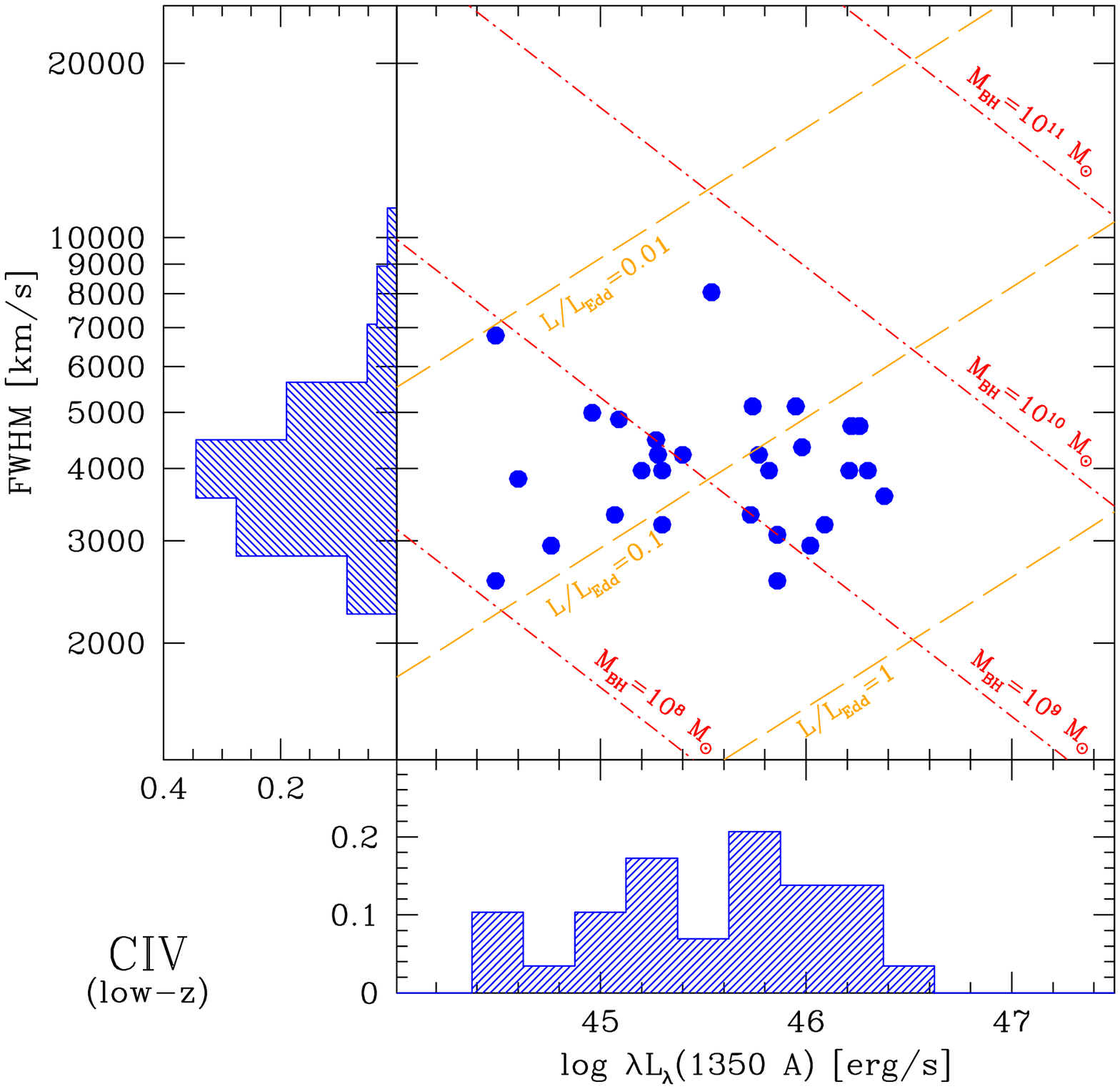}\\
\includegraphics[width=0.4\textwidth]{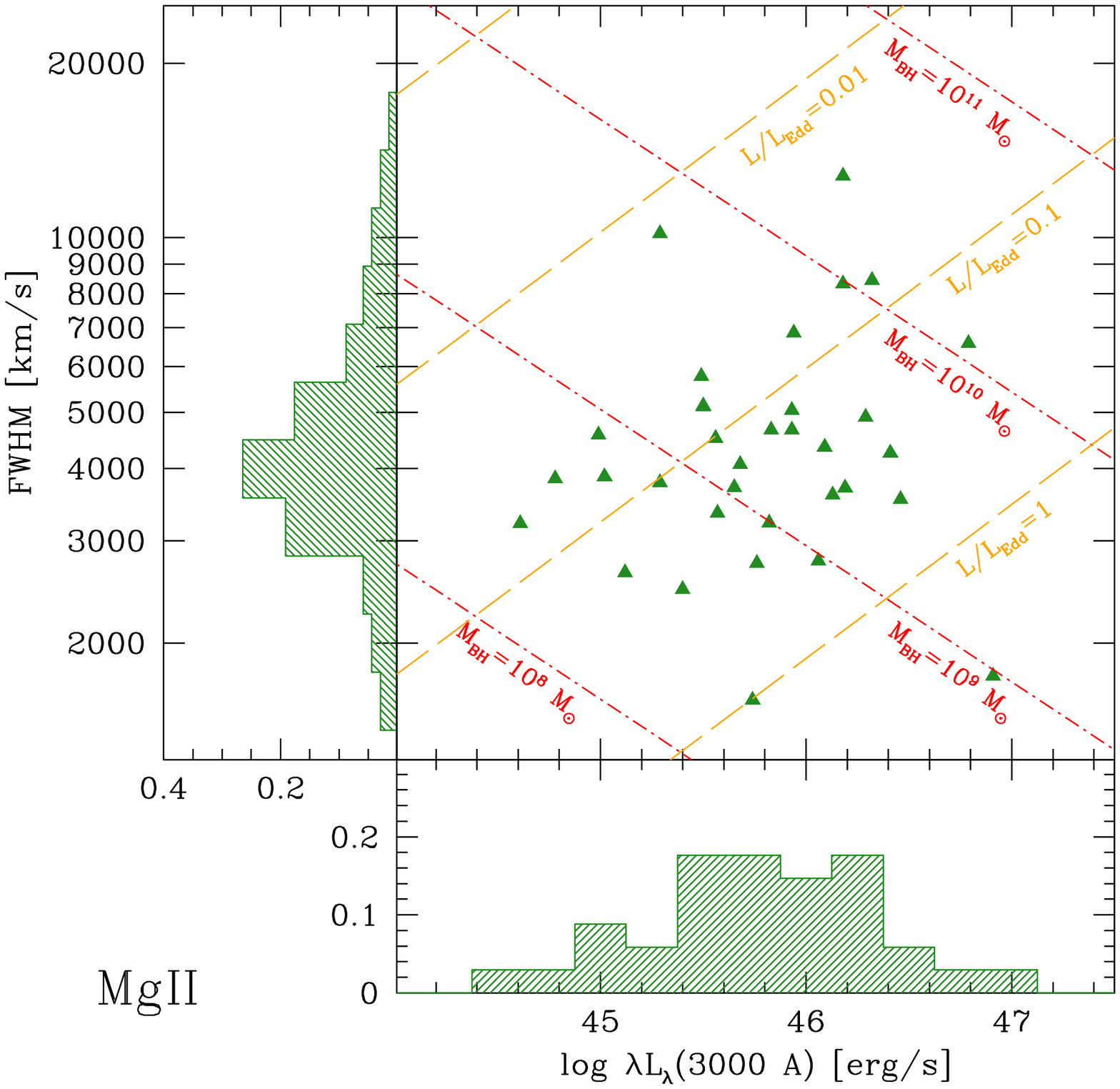}
\includegraphics[width=0.4\textwidth]{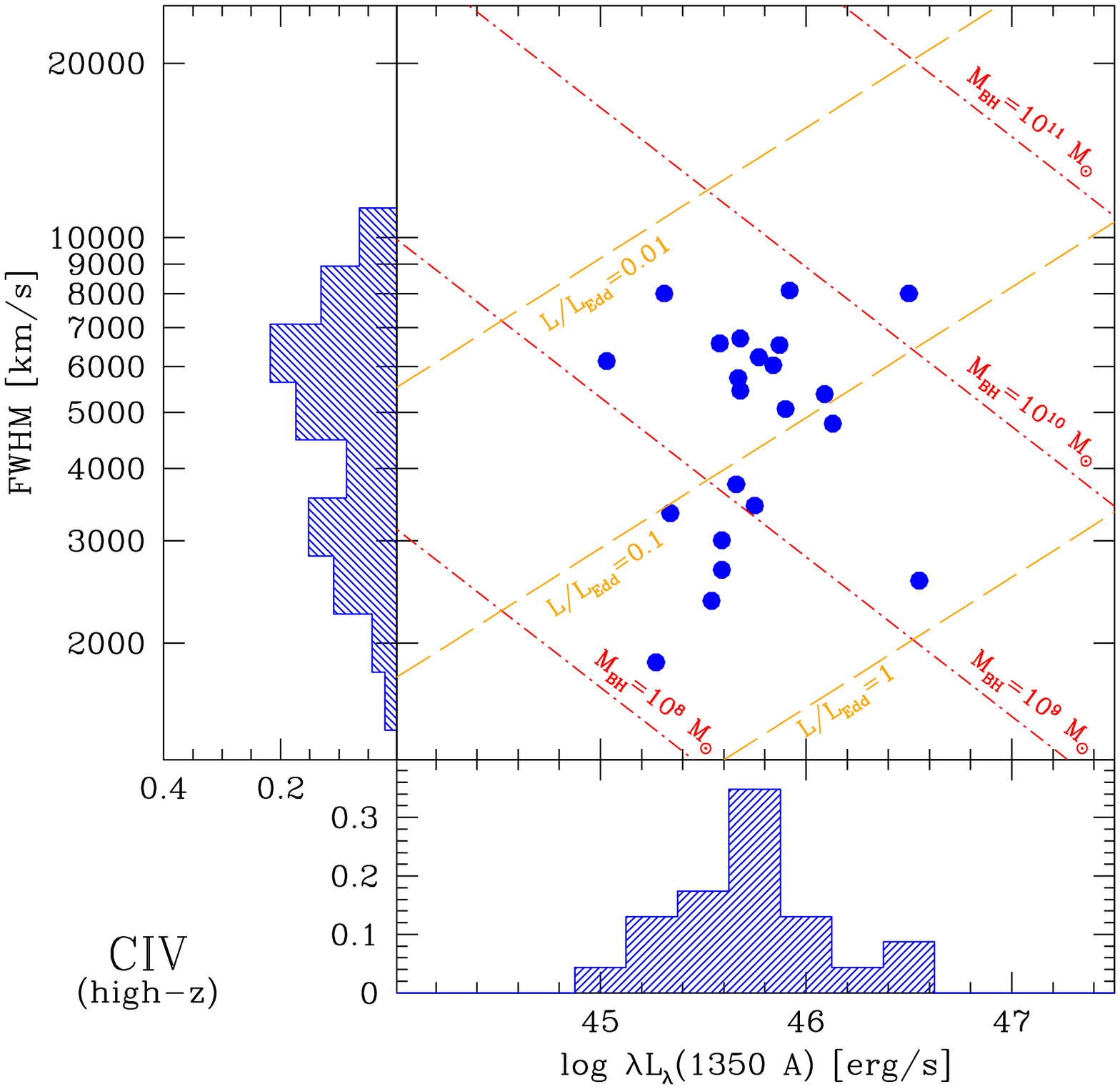}\\
\caption{The distribution of our data in the (\lLl,FWHM) plane, for
\Hb{}, \Civ{} at low-$z$ (upper panels), \Mgii{} and \Civ{} at high-$z$
(bottom panels). The (normalized) projections along each axis are
also shown. The loci with constant \Mbh{} (dot-dashed lines) and \Edd{}
(dashed lines) are plotted. Concerning the luminosities, low-$z$ \Civ{}
data show on average
The values of
\Hb{} FWHM show a wider distribution with respect to \Mgii{}. data are log-normally distributed
both in the \lLl{} and in the FWHM values. \Civ{} data show less spread
in the sampled \lLl{} space, and a flatter distribution in the line width
with respect to the \Mgii{}.
}\label{fig_labita}
\end{center}
\end{figure*}
Concerning the fitted function, a single gaussian usually does not
provide satisfactory fits, especially for \Civ{}
\citep[see][]{decarli08a,shen08,bon09}.
We performed our analysis both with the sum of 2 gaussian profiles with the
same peak (hereafter, G2) and the Gauss-Hermite series \citep[GH;][]{vandermarel93},
truncated at the fourth order. Reduced-$\chi^2$ values obtained from these
functions are similarly good, suggesting that both functions can describe
line profiles reasonably well.
No significant offset is observed among the line width estimates based on the
two functions, the residuals lying within 20 per cent
\citep[see also Figure A1 in][]{decarli08a}.
We thus conclude that the two functions are equivalent to our purposes.
We note that, since G2 fits are necessarily symmetric, while GH fits are
not, the consistency between the two estimates of the line width suggests
that the FWHM is poorly sensitive to line profile asymmetries.
Throughout the paper, we will refer to the FWHM estimates based
on GH fits.

Summarizing, we estimate that the typical uncertainties in the line width
estimates due to the adopted fit procedures lie within 20 per cent, and
usually even lower (if some modelling of the \FeII{} emission is adopted).
Reduced-$\chi^2$ maps suggest that, given the high signal to noise ratio of
our data (exceeding 20 in all but few cases), formal uncertainties in the
parameter estimates contribute to $\lsim 10$ per cent of the FWHM value.
Therefore, we conclude that typical uncertainties in the FWHM estimates
are around 20 per cent.

\section[]{Virial estimates of BH masses}\label{sec_results}

In this section, we match the new estimates of the continuum luminosity,
FWHM and \Mbh{} with those from our previous low-$z$ studies
\citep{labita06,decarli08a}.

In Figure \ref{fig_labita}, we plot the distribution of our data in
the (\lLl,FWHM) plane \citep[see a similar approach
in][]{fine08,labita09a}. This allows us to monitor how observable
quantities (here, the continuum luminosity of the quasar and the
line widths) affect our estimates of \Mbh{} and the Eddington ratio,
\Edd, derived \emph{assuming} equation \ref{eq_virial}, $f({\rm
H\beta,MgII})=1.6$, $f({\rm CIV})=2.4$ \citep[as defined in equation
\ref{eq_def_f} and in][]{decarli08a} and the bolometric corrections
given in \citet{richards06}: $L_{\rm bol}/$\lLl{}=$9.26$, $5.15$ and
$3.81$ for \Hb{}, \Mgii{} and \Civ{} respectively. Average and rms
values of FWHM, \lLl{}, \Mbh{} and \Edd{} are provided in Table
\ref{tab_ave_results}, while data of individual quasars are given in
table \ref{tab_fits}.

The monochromatic luminosity at 5100 \AA{} is on average $3.5$ times
fainter than at 1350 \AA{} in the same redshift bin. This is consistent
with an average power-law index of the quasar continuum $\alpha$ close to 2
(defined so that $F_\lambda\propto\lambda^{-\alpha}$).
The 3000 \AA{} luminosities sampled in our study range from few times
$10^{44}$ to $10^{47}$ erg/s, the bulk being around $10^{46}$ erg/s. The
\Civ{} data at high-$z$ are more clustered around approximately the same
luminosity as \Mgii{} data, $\approx 6\cdot 10^{45}$ erg/s. This thinner
distribution is due to the combination of two selection effects: lowest
luminosity quasars
are missed due to the Malmquist bias (the \Civ{} line falls in the optical
bands at $z>1.6$, see Figure \ref{fig_distr_z}); highest luminosity quasars
were rejected in the studies of the host galaxy luminosities, in order to make
the detection of the extend emission around the nuclear source more feasible
\citep[see Figure 1 and ][]{kotilainen09}.

Concerning the line widths, \Hb{} values show a wider dispersion with respect
to the \Mgii{} line, notwithstanding the two lines have similar ionization
potential, thus they are believed to originate in the same regions
\citep[see also][]{shen08,labita09a,labita09b}. Low-redshift data from the \Civ{} line show
a smaller average value with respect to both \Hb{} and high-$z$ \Civ{} data,
and a significantly narrower distribution \citep[see the discussion in][]{decarli08a}.

With only few exceptions, all our data reside in the
locus defined by $10^{8} \gsim$ \Mbh{}/\Msun{} $\gsim 10^{10}$ and
$0.01 \gsim$ \Edd{} $\gsim 1$.

\begin{table}
\begin{center}
\caption{Average and rms values of FWHM (2), \lLl{} (3), \Mbh{} (4) and
\Edd{} (5) for all the subsamples in this study.
} \label{tab_ave_results}
\begin{tabular}{ccccc}
   \hline
   Line             &$\langle$log FWHM$\rangle$&$\langle$log \lLl$\rangle$&$\langle$log \Mbh{}$\rangle$&$\langle$log \Edd{}$\rangle$ \\
                    &  [km/s]          & [erg/s]          & [\Msun]          &                  \\
   (1)              &  (2)             &  (3)             & (4)              & (5)              \\
   \hline
   \multicolumn{5}{l}{{\it Low-$z$}} \\
   \Hb{}            & $3.66 \pm 0.19$ & $45.01 \pm 0.51$ & $9.01 \pm 0.47$ & $-1.13 \pm 0.45$ \\
   \Civ{}           & $3.60 \pm 0.11$ & $45.55 \pm 0.56$ & $9.06 \pm 0.38$ & $-1.03 \pm 0.34$ \\
   \hline
   \multicolumn{5}{l}{{\it High-$z$}} \\
   \Mgii{}          & $3.63 \pm 0.19$ & $45.78 \pm 0.54$ & $9.21 \pm 0.49$ & $-0.82 \pm 0.46$ \\
   \Civ{}           & $3.66 \pm 0.19$ & $45.74 \pm 0.36$ & $9.28 \pm 0.47$ & $-1.06 \pm 0.41$ \\
   \hline
   \end{tabular}
   \end{center}
\end{table}



\begin{figure}
\begin{center}
\includegraphics[width=0.49\textwidth]{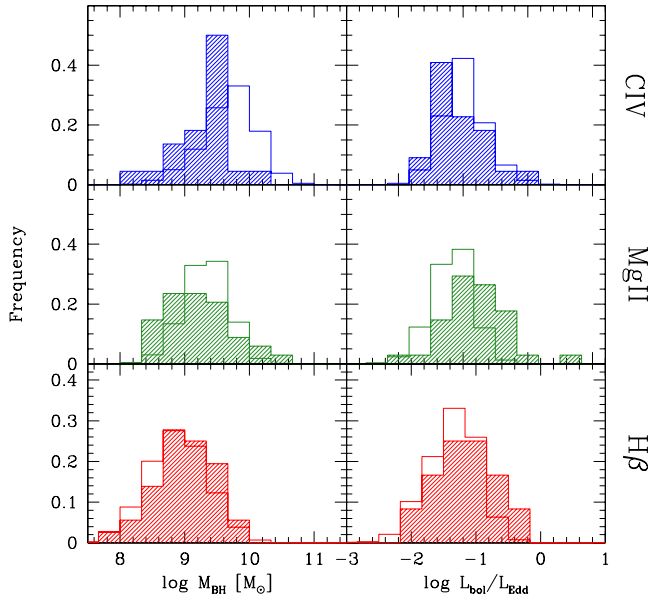}\\
\caption{ The distributions of \Mbh{} and \Edd{} as derived from
\Civ{} (\emph{upper panels}), \Mgii{} (\emph{central panels})
and \Hb{} (\emph{bottom panels}). Low-$z$ \Civ{}-based estimates
are not included, as no counter-part is available in the \citet{shen08}
sample. Shaded histograms refer to our objects, while empty histograms
refer to the dataset of \citet{shen08}.
}\label{fig_shen}
\end{center}
\end{figure}

In Figure \ref{fig_shen} we compare the distributions of \Mbh{} and \Edd{}
from our dataset with the huge quasar sample from
\citet[][S08]{shen08}\footnote{From Shen et al. (2008) we take the values
of $z$, FWHM and \lLl{} of the objects in the same redshift ranges than our
study. All the derived quantities (e.g., \Mbh{} and \Edd{}) are re-computed
following the recipes discussed in this work.}. We dropped our low-$z$,
\Civ{}-based data from this comparison as they do not have any counter-part
in the S08 data. The distributions of black
hole masses and Eddington ratios computed from \Hb{} in our study are in
substantial agreement with those by S08. On the other hand, the black
hole masses in our sample have increasingly smaller values than those
from S08. We interpret this trend as the superposition of a number of
effects: the luminosity cut adopted by \citet{kotilainen09} for $z>2$ objects,
the evolution of the luminosity and mass functions of active black holes
through redshift and the occurrence of the Malmquist bias
\citep[on this topic, see][]{labita09a,labita09b}. On average, the Eddington ratios sampled
in our study do not show significant differences with respect to those in S08,
nor evolution from $z=0$ to $z=3$.

\begin{figure}
\begin{center}
\includegraphics[angle=-90, width=0.49\textwidth]{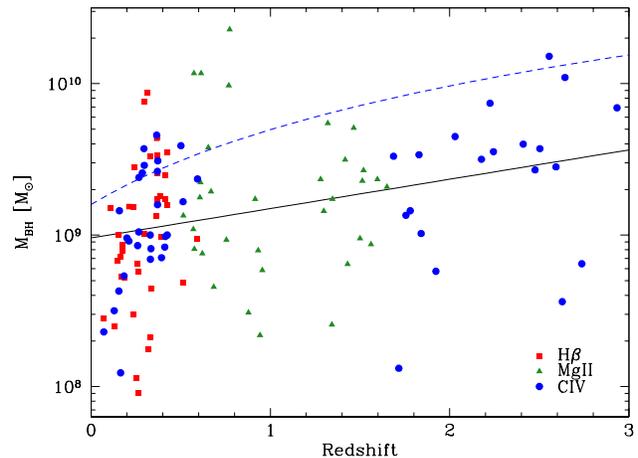}\\
\caption{The redshift dependence of \Mbh{} for the quasars in our
sample. Circles, triangles and squares refer to \Civ{}, \Mgii{}
and \Hb{}-based \Mbh{} estimates. The best linear fit is shown as
a solid line. The dashed line refers to the redshift evolution of the
maximum \Mbh{} as derived by \citet{labita09b}.
}\label{fig_mbh_z}
\end{center}
\end{figure}

The relatively small statistics and the luminosity-based selection of the
targets in our sample hinder the study of the dependence of \Mbh{} on the
redshift. We note that, on average, the higher is the redshift, the more
massive are the black holes (see Figure \ref{fig_mbh_z}). The linear best
fit is:
\begin{equation}
\log {\cal M}_{\rm BH}/{\rm M}_{\odot} = (0.19 \pm 0.06) z + (8.98 \pm 0.06)
\end{equation}
This trend is in qualitative agreement with the one found by
\citet{labita09b} with a procedure aimed to minimize the Malmquist bias.
However, the occurrence of selection biases cannot be ruled
out in the present work.

\section{Conclusions}\label{sec_conclusions}

We start from a list of 96 quasars with known host galaxy
luminosity. New, high-quality (S/N$\sim$30) spectra of the mid- and
high-redshift targets ($z>0.5$) are presented here and matched with
those of low-$z$ quasars we collected in our previous studies. We
analyse the continuum luminosity and profiles of broad emission
lines in order to infer black hole masses. In particular, the \Civ{}
line is studied both at low and high redshift, thus avoiding a
systematic error related to the adopted emission line in the
estimate of \Mbh{}. We found that high redshift quasars in our
sample do have, on average, larger BH masses than local ones, but we
note that this result is potentially affected by the luminosity
selection and the Malmquist bias. In a accompanying  Paper II, we
study the ratio between \Mbh{} and the luminosity and mass of the
host galaxies of the quasars in our sample, and its evolution as a
function of the redshift.

\section*{Acknowledgments}
We thank the anonymous referee for his/her fruitful suggestions.
This work was based on observations
made with the Nordic Optical Telescope (programme ID:
35-039 \& 36-039), operated
on the island of La Palma jointly by Denmark, Finland, Iceland,
Norway, and Sweden, in the Spanish Observatorio del Roque de los
Muchachos of the Instituto de Astrofisica de Canarias, and with
the European Southern Observatory 3.6m telescope in La Silla
[programme ID: 078.B-0087(A) \& 079.B-0304(A)].
ALFOSC \@ NOT is owned by the Instituto de Astrofisica de Andalucia
(IAA) and operated at the Nordic Optical Telescope under agreement
between IAA and the NBIfAFG of the Astronomical Observatory of Copenhagen.

This research has made use of the \emph{VizieR Service}, available
at \texttt{http://vizier.u-strasbg.fr/viz-bin/VizieR} and of the
NASA/IPAC Extragalactic Database (NED) which is operated by the Jet
Propulsion Laboratory, California Institute of Technology, under
contract with the National Aeronautics and Space Administration.
Observed spectra are available at \verb|www.dfm.uninsubria.it/astro/caqos/|.

\appendix
\section[]{Notes on individual targets}\label{sec_appendix}

The catalogue by \citet{vcv06} reports tentative redshifts for quasars 0119-370
\citep[$z=1.32$,][]{savage84} and 0152-4055 \citep[$z=1.65$,][]{drinkwater87}.
Our spectra confirm both them: from the peak wavelength of the \Mgii{} line, we infer
$z=1.369$ for 0119-370, and $z=1.632$ for 0152-4055.

PKS 0440-00 has a literature redshift of $0.844$, inferred from
the detection of a possible emission line at 5158 \AA{}
\citep{schmidt77}. Our spectrum clearly shows that the \Mgii{} line
is actually present, but it is peaked at 4496 \AA{}, yielding $z=0.607$.
The \FeII{} features around \Mgii{} support the line identification.
In particular, we note that a bump typical of the \FeII{} system red-wards
the \Mgii{} line is observed at $\sim 5150$ \AA{}.


ZC2351+010B was classified as a quasar in objective prism survey by
\citet{zhan89}, with a tentative redshift $z=0.810$. Our spectrum
shows the typical features of a M-type star \citep[see a similar case in][]{decarli09b}.

Quasars 4C02.54 and PKS1102-242 were observed during N36 and E78 respectively.
Since the broad lines of interest fall right on the edges of the observed ranges of
the collected spectra, we can not infer reliable estimates of \Mbh{} for these objects.

\label{lastpage}


\onecolumn
\begin{longtable}{@{}ccccccccc@{}}
  \caption{ The observed sample. All catalogue values are referred to \citet{vcv06}.
  (1) quasar name; (2),(3) quasar right ascension and declination in J2000; (4) catalogue
  redshift; (5) radio loudness (L=radio loud, Q=radio quiet); (6) catalogue apparent V-band
  magnitude; (7) Galactic V-band extinction; (8) absolute restframe V-band
  magnitude, as derived from the catalogue apparent V-band magnitude, $k$-corrected
  assuming the \citet{francis91} template; (9) data sources: L06=\citet{labita06};
    D08=\citet{decarli08a}; E$xx$=new ESO/3.6m observations (E77=September, 2006;
    E78=March, 2007; E79=September, 2007); N$xx$=new NOT observations
    (N35=April, 2007; N36=October, 2007).}\label{tab_sample}\\
\hline\\[-2ex]
   Name         &  RA(J2000) & dec (J2000)& $z$ & Radio & $m_{\rm V}$ & A$_{\rm V}$ & $M_{\rm V}$ & Ref \\
                &            &           &   & loudness  & [mag] & [mag] & [mag]&     \\
   (1)          &   (2)      & (3)       & (4) & (5)   & (6)   & (7)   & (8)    & (9)  \\
\hline
\endfirsthead
\hline\\[-2ex]
   Name         &  RA(J2000) & dec (J2000)& $z$ & Radio & $m_{\rm V}$ & A$_{\rm V}$ & $M_{\rm V}$ & Ref \\
                &            &           &   & loudness  & [mag] & [mag] & [mag]&     \\
   (1)          &   (2)      & (3)       & (4) & (5)   & (6)   & (7)   & (8)    & (9)  \\
\hline
\endhead
\\[-1.8ex] \hline
\multicolumn{9}{l}{\it Continue in next page}\\
\endfoot
\\[-1.8ex] \hline
\endlastfoot
         PKS0000-177 & 00:03:22.0 &  -17:27:12 & 1.465 & L & 19.00 & 0.086 & -26.39 &     E77 \\
          Q0040-3731 & 00:42:57.4 &  -37:15:31 & 1.780 & Q & 17.80 & 0.053 & -28.15 &     E77 \\
             SGP2:36 & 00:51:14.3 &  -29:05:20 & 1.756 & Q & 19.62 & 0.046 & -26.29 &     E79 \\
             SGP5:46 & 00:52:22.8 &  -27:30:03 & 0.955 & Q & 18.88 & 0.048 & -25.38 &     E79 \\
            0054+144 & 00:57:09.9 &  +14:46:10 & 0.171 & Q & 15.70 & 0.151 & -24.09 &     D08 \\
             SGP4:39 & 00:59:08.9 &  -27:51:25 & 1.716 & Q & 19.64 & 0.074 & -26.21 &     E79 \\
         PKS0100-270 & 01:02:56.3 &  -26:46:36 & 1.597 & L & 17.80 & 0.051 & -27.83 &     E77 \\
       LBQS0100+0205 & 01:03:13.0 &  +02:21:10 & 0.393 & Q & 17.51 & 0.069 & -24.19 &     L06 \\
            0110+297 & 01:13:24.2 &  +29:58:15 & 0.363 & L & 17.00 & 0.209 & -24.50 &     D08 \\
         PKS0113-283 & 01:15:23.9 &  -28:04:55 & 2.555 & L & 17.10 & 0.051 & -29.71 &     E79 \\
            0119-370 & 01:21:24.1 &  -36:50:02 & 1.320 & Q & 19.20 & 0.058 & -25.92 &     E79 \\
            0133+207 & 01:36:24.4 &  +20:57:27 & 0.425 & L & 18.10 & 0.203 & -23.80 & L06,D08 \\
                3C48 & 01:37:41.3 &  +33:09:35 & 0.367 & L & 16.20 & 0.144 & -25.31 & L06,D08 \\
        HB890137+012 & 01:39:57.2 &  +01:31:46 & 0.260 & L & 17.10 & 0.096 & -23.63 &     L06 \\
           0152-4055 & 01:54:20.1 &  -40:40:30 & 1.650 & Q & 19.30 & 0.055 & -26.44 &     E79 \\
         PKS0155-495 & 01:57:38.0 &  -49:15:19 & 1.298 & L & 18.40 & 0.053 & -26.68 &     E77 \\
          PKS0159-11 & 02:01:57.1 &  -11:32:34 & 0.669 & L & 16.40 & 0.073 & -26.77 &     N36 \\
          B0204+2916 & 02:07:02.2 &  +29:30:46 & 0.109 & Q & 16.80 & 0.212 & -21.85 &     D08 \\
            0244+194 & 02:47:40.8 &  +19:40:58 & 0.176 & Q & 16.66 & 0.368 & -23.14 &     D08 \\
       KUV03086-0447 & 03:11:04.7 &  -04:35:41 & 0.755 & Q & 17.50 & 0.284 & -26.05 &     N36 \\
            MZZ01558 & 03:14:51.5 &  -54:57:14 & 1.829 & Q & 21.64 & 0.054 & -24.37 &     E79 \\
              US3828 & 03:18:25.6 &  +15:59:56 & 0.515 & Q & 16.90 & 0.395 & -25.53 &     N36 \\
          Q0335-3546 & 03:37:02.4 &  -35:36:39 & 1.841 & Q & 19.80 & 0.043 & -26.23 &     E79 \\
         PKS0348-120 & 03:51:11.0 &  -11:53:23 & 1.520 & L & 19.00 & 0.169 & -26.49 &     E77 \\
          PKS0349-14 & 03:51:28.6 &  -14:29:10 & 0.614 & L & 16.20 & 0.263 & -26.74 &     N36 \\
         PKS0402-362 & 04:03:53.8 &  -36:05:02 & 1.417 & L & 17.17 & 0.017 & -28.13 &     E79 \\
         PKS0403-132 & 04:05:34.0 &  -13:08:14 & 0.571 & L & 17.10 & 0.192 & -25.64 &     N36 \\
         PKS0405-123 & 04:07:48.5 &  -12:11:36 & 0.574 & L & 14.90 & 0.191 & -27.85 &     N36 \\
          PKS0414-06 & 04:17:16.7 &  -05:53:45 & 0.773 & L & 15.90 & 0.143 & -27.72 &     N36 \\
         PKS0420-014 & 04:23:15.8 &  -01:20:33 & 0.915 & L & 17.00 & 0.436 & -27.14 &     N36 \\
          PKS0440-00 & 04:42:38.6 &  -00:17:43 & 0.607 & L & 18.41 & 0.174 & -24.49 &     N36 \\
           0624+6907 & 06:30:02.5 &  +69:05:04 & 0.370 & Q & 14.20 & 0.324 & -27.36 & L06,D08 \\
          PKS0710+11 & 07:13:02.3 &  +11:46:15 & 0.768 & L & 16.60 & 0.488 & -27.00 &     N36 \\
       MS0824.2+0327 & 08:26:52.9 &  +03:17:13 & 1.431 & Q & 20.20 & 0.132 & -25.13 &     E78 \\
        MS08287+6614 & 08:33:17.9 &  +66:03:46 & 0.610 & Q & 18.00 & 0.131 & -24.93 &     N36 \\
          PKS0838+13 & 08:40:47.6 &  +13:12:23 & 0.684 & L & 18.10 & 0.309 & -25.15 &     N35 \\
              US1867 & 08:53:34.2 &  +43:49:02 & 0.513 & Q & 16.40 & 0.112 & -26.02 & L06,D08 \\
            0903+169 & 09:06:31.9 &  +16:46:11 & 0.411 & L & 18.27 & 0.133 & -23.56 & L06,D08 \\
              TON392 & 09:12:17.8 &  +24:50:38 & 0.654 & Q & 16.00 & 0.148 & -27.12 &     N36 \\
        MS09441+1333 & 09:46:52.0 &  +13:20:26 & 0.131 & Q & 16.05 & 0.132 & -23.15 &     D08 \\
            0953+415 & 09:56:52.4 &  +41:15:22 & 0.234 & Q & 15.30 & 0.042 & -25.17 &     D08 \\
            1001+291 & 10:04:02.5 &  +28:55:35 & 0.330 & Q & 15.50 & 0.072 & -25.79 & L06,D08 \\
            1004+130 & 10:07:26.1 &  +12:48:56 & 0.240 & L & 15.20 & 0.127 & -25.35 &     D08 \\
        Z101733-0203 & 10:17:33.5 &  -02:03:07 & 1.343 & Q & 20.80 & 0.145 & -24.37 &     E78 \\
          PKS1015-31 & 10:18:09.3 &  -31:44:14 & 1.346 & L & 20.40 & 0.277 & -24.77 &     E78 \\
          PKS1018-42 & 10:20:03.9 &  -42:51:30 & 1.280 & L & 18.80 & 0.380 & -26.26 &     E78 \\
            1058+110 & 11:00:47.8 &  +10:46:13 & 0.423 & L & 17.10 & 0.085 & -24.80 &     D08 \\
            1100+772 & 11:04:13.7 &  +76:58:58 & 0.315 & L & 15.72 & 0.112 & -25.43 &     D08 \\
            1116+215 & 11:19:27.6 &  +21:17:20 & 0.177 & Q & 14.70 & 0.075 & -25.10 &     D08 \\
            1150+497 & 11:53:24.4 &  +49:31:09 & 0.334 & L & 17.10 & 0.071 & -24.20 & L06,D08 \\
            1202+281 & 12:04:42.1 &  +27:54:11 & 0.165 & Q & 15.60 & 0.070 & -24.06 & L06,D08 \\
            1208+322 & 12:10:37.6 &  +31:57:06 & 0.388 & L & 16.00 & 0.056 & -25.64 &     D08 \\
            1216+069 & 12:19:20.9 &  +06:38:38 & 0.331 & Q & 15.65 & 0.072 & -25.64 &     L06 \\
             MRK0205 & 12:21:44.0 &  +75:18:38 & 0.071 & Q & 15.24 & 0.139 & -22.65 & L06,D08 \\
            1222+125 & 12:25:12.9 &  +12:18:36 & 0.415 & L & 17.86 & 0.115 & -23.98 &     D08 \\
               3C273 & 12:29:06.7 &  +02:03:08 & 0.158 & L & 12.90 & 0.068 & -26.63 &     L06 \\
            1230+097 & 12:33:25.8 &  +09:31:23 & 0.415 & Q & 16.15 & 0.068 & -25.69 & L06,D08 \\
        Z124029-0010 & 12:40:29.7 &  -00:10:48 & 2.030 & Q & 19.80 & 0.074 & -26.45 &     E78 \\
          PG1302-102 & 13:05:33.0 &  -10:33:19 & 0.286 & L & 15.20 & 0.141 & -25.71 &     L06 \\
            1307+085 & 13:09:47.0 &  +08:19:49 & 0.155 & Q & 15.10 & 0.112 & -24.43 & L06,D08 \\
            1309+355 & 13:12:17.8 &  +35:15:21 & 0.184 & L & 15.64 & 0.040 & -24.29 & L06,D08 \\
        Z133136-0002 & 13:31:36.2 &  -00:02:53 & 2.710 & Q & 20.70 & 0.084 & -26.22 &     E78 \\
            1402+436 & 14:04:38.8 &  +43:27:07 & 0.320 & Q & 15.62 & 0.035 & -25.60 &     D08 \\
          PG1416-129 & 14:19:05.7 &  -13:10:56 & 0.129 & Q & 16.10 & 0.311 & -22.94 &     L06 \\
            1425+267 & 14:27:35.5 &  +26:32:14 & 0.366 & L & 15.68 & 0.062 & -25.83 & L06,D08 \\
        Z143220-0215 & 14:32:20.1 &  -02:15:47 & 2.476 & Q & 20.40 & 0.148 & -26.35 &     E78 \\
        Z144022-0122 & 14:40:22.3 &  -01:22:33 & 2.244 & Q & 20.00 & 0.166 & -26.51 &     E78 \\
            1444+407 & 14:46:45.9 &  +40:35:06 & 0.267 & Q & 15.70 & 0.046 & -25.04 & L06,D08 \\
         PKSJ1511-10 & 15:13:44.9 &  -10:12:00 & 1.513 & L & 18.80 & 0.351 & -26.68 &     E78 \\
             1512+37 & 15:14:43.0 &  +36:50:50 & 0.371 & L & 16.27 & 0.072 & -25.30 & L06,D08 \\
          PKS1524-13 & 15:26:59.4 &  -13:51:01 & 1.687 & L & 20.50 & 0.400 & -25.29 &     E78 \\
             3C323.1 & 15:47:43.5 &  +20:52:17 & 0.266 & L & 16.70 & 0.140 & -24.04 & L06,D08 \\
            1549+203 & 15:52:02.3 &  +20:14:02 & 0.250 & Q & 16.40 & 0.176 & -24.25 &     D08 \\
         HS1623+7313 & 16:22:16.8 &  +73:06:15 & 0.621 & Q & 16.30 & 0.110 & -26.68 &     N35 \\
            1635+119 & 16:37:46.5 &  +11:49:50 & 0.146 & Q & 16.50 & 0.171 & -22.88 &     D08 \\
               3C345 & 16:42:58.8 &  +39:48:37 & 0.594 & L & 15.96 & 0.044 & -26.89 & L06,D08 \\
               3C351 & 17:04:41.4 &  +60:44:31 & 0.372 & L & 15.28 & 0.075 & -26.29 & L06,D08 \\
            1821+643 & 18:21:57.3 &  +64:20:36 & 0.297 & Q & 14.10 & 0.141 & -26.89 & L06,D08 \\
               3C422 & 20:47:10.4 &  -02:36:23 & 0.942 & L & 18.69 & 0.181 & -25.54 &     N36 \\
          MC2112+172 & 21:14:56.7 &  +17:29:23 & 0.878 & L & 17.90 & 0.446 & -26.12 &     N36 \\
          Q2125-4432 & 21:29:01.0 &  -44:19:50 & 2.503 & Q & 20.39 & 0.077 & -26.38 &     E79 \\
          PKS2128-12 & 21:31:35.3 &  -12:07:04 & 0.501 & L & 16.11 & 0.204 & -26.26 &     L06 \\
          PKS2135-14 & 21:37:45.2 &  -14:32:55 & 0.200 & L & 15.50 & 0.169 & -24.65 &     L06 \\
            2141+175 & 21:43:35.5 &  +17:43:49 & 0.211 & L & 15.73 & 0.367 & -24.53 & L06,D08 \\
        Z215539-3026 & 21:55:39.7 &  -30:26:23 & 2.593 & Q & 20.44 & 0.085 & -26.39 &     E79 \\
            2201+315 & 22:03:15.0 &  +31:45:38 & 0.295 & L & 15.58 & 0.410 & -25.40 & L06,D08 \\
          PKS2204-20 & 22:07:33.9 &  -20:38:35 & 1.923 & L & 19.49 & 0.098 & -26.63 &     E79 \\
        Z221139-3132 & 22:11:39.1 &  -31:32:53 & 2.391 & Q & 20.40 & 0.048 & -26.28 &     E79 \\
        Z222702-3205 & 22:27:02.4 &  -32:05:36 & 2.177 & Q & 20.13 & 0.044 & -26.29 &     E79 \\
          Q2225-403A & 22:28:49.9 &  -40:08:34 & 2.410 & Q & 20.20 & 0.042 & -26.50 &     E79 \\
          Q2225-403B & 22:28:50.4 &  -40:08:27 & 0.932 & Q &  0.00 & 0.042 & -44.20 &     E79 \\
          PKS2227-08 & 22:29:40.1 &  -08:32:54 & 1.562 & L & 17.50 & 0.170 & -28.06 &     E77 \\
        Z223048-2954 & 22:30:48.1 &  -29:54:05 & 2.652 & Q & 20.56 & 0.051 & -26.32 &     E79 \\
            2247+140 & 22:50:25.3 &  +14:19:52 & 0.235 & L & 16.93 & 0.168 & -23.54 &     D08 \\
        Z225950-3206 & 22:59:50.8 &  -32:06:03 & 2.225 & Q & 19.72 & 0.066 & -26.77 &     E79 \\
        Z231751-3147 & 23:17:51.9 &  -31:47:40 & 2.628 & Q & 20.58 & 0.053 & -26.28 &     E79 \\
        Z232755-3154 & 23:27:55.5 &  -31:54:36 & 2.737 & Q & 20.73 & 0.053 & -26.20 &     E79 \\
        Z233451-2929 & 23:34:52.0 &  -29:29:20 & 2.669 & Q & 20.72 & 0.060 & -26.17 &     E79 \\
         PKS2345-167 & 23:48:02.6 &  -16:31:13 & 0.576 & L & 18.40 & 0.086 & -24.35 &     N36 \\
          Q2348-4012 & 23:51:02.1 &  -39:56:18 & 1.500 & Q & 19.50 & 0.047 & -25.96 &     E79 \\
\end{longtable}

\newpage

\begin{longtable}{@{}ccccccc@{}}
\caption{Journal of the new observations. (1) quasar name; (2) catalogue redshift;
  (3) run of observation (see table \ref{tab_runs} for the description of each run);
  (4) date of observations; (5) seeing, computed from the profiles of field stars
  in corollary images; (6) spectral signal to noise ratio (per pixel); (7)
  uncertainty in the photometric calibration (in magnitudes).}\label{tab_journal}\\
 \hline\\[-2ex]
   Name                &  $z$  & Run & Obs.Date & Seeing   &  S/N   & $\delta$ZP  \\
                       &       &     &[dd/mm/yy]& [arcsec] &        &[mag] \\
   (1)                 &   (2) & (3) & (4)      & (5)      & (6)    & (7)  \\
\hline
\endfirsthead
 \hline\\[-2ex]
   Name                &  $z$  & Run & Obs.Date & Seeing   &  S/N   & $\delta$ZP  \\
                       &       &     &[dd/mm/yy]& [arcsec] &        &[mag] \\
   (1)                 &   (2) & (3) & (4)      & (5)      & (6)    & (7)  \\
\hline
\endhead
\\[-1.8ex] \hline
\multicolumn{7}{l}{\it Continue in next page}\\
\endfoot
\\[-1.8ex] \hline
\endlastfoot
          PKS0000-177  & 1.465 & E77 & 30/09/05 & $ 1.9 $ &   17.8 & 0.02 \\
           Q0040-3731  & 1.780 & E77 & 30/09/05 & $ 2.0 $ &   19.5 & 0.02 \\
              SGP2:36  & 1.756 & E79 & 10/09/07 & $ 1.2 $ &   13.3 & 0.11 \\
              SGP5:46  & 0.955 & E79 & 10/09/07 & $ 1.5 $ &   26.1 & 0.06 \\
              SGP4:39  & 1.716 & E79 & 11/09/07 & $ 1.4 $ &   38.4 & 0.05 \\
          PKS0100-270  & 1.597 & E77 & 30/09/05 & $ 1.7 $ &   17.5 & 0.02 \\
          PKS0113-283  & 2.555 & E79 & 11/09/07 & $ 1.5 $ &   26.2 & 0.09 \\
             0119-370  & 1.320 & E79 & 11/09/07 & $ 1.4 $ &   31.7 & 0.11 \\
            0152-4055  & 1.650 & E79 & 12/09/07 & $ 1.5 $ &   20.7 & 0.04 \\
          PKS0155-495  & 1.298 & E77 & 30/09/05 & $ 1.8 $ &   33.9 & 0.02 \\
           PKS0159-11  & 0.669 & N36 & 17/10/07 & $ 0.7 $ &   63.4 & 0.07 \\
               PB6708  & 0.868 & N36 & 17/10/07 & $ 0.8 $ &   34.7 & 0.07 \\
        KUV03086-0447  & 0.755 & N36 & 19/10/07 & $ 1.5 $ &   24.0 & 0.03 \\
             MZZ01558  & 1.829 & E79 & 12/09/07 & $ 2.0 $ &   26.9 & 0.12 \\
               US3828  & 0.515 & N36 & 18/10/07 & $ 0.7 $ &    9.1 & 0.02 \\
           Q0335-3546  & 1.841 & E79 & 09/09/07 & $ 1.0 $ &   25.2 & 0.10 \\
          PKS0348-120  & 1.520 & E77 & 30/09/05 & $ 1.7 $ &  103.7 & 0.02 \\
           PKS0349-14  & 0.614 & N36 & 17/10/07 & $ 0.8 $ &    6.0 & 0.07 \\
          PKS0402-362  & 1.417 & E79 & 09/09/07 & $ 1.0 $ &   97.3 & 0.15 \\
          PKS0403-132  & 0.571 & N36 & 19/10/07 & $ 1.3 $ &   17.5 & 0.03 \\
          PKS0405-123  & 0.574 & N36 & 17/10/07 & $ 0.8 $ &   21.7 & 0.07 \\
           PKS0414-06  & 0.773 & N36 & 17/10/07 & $ 0.6 $ &   15.4 & 0.07 \\
          PKS0420-014  & 0.915 & N36 & 19/10/07 & $ 1.1 $ &   10.5 & 0.01 \\
           PKS0440-00  & 0.844 & N36 & 18/10/07 & $ 0.7 $ &   30.8 & 0.00 \\
           PKS0710+11  & 0.768 & N36 & 19/10/07 & $ 1.1 $ &   70.6 & 0.03 \\
        MS0824.2+0327  & 1.431 & E78 & 23/03/07 & $ 1.0 $ &   13.4 & 0.05 \\
         MS08287+6614  & 0.610 & N36 & 18/10/07 & $ 0.9 $ &   33.7 & 0.02 \\
           PKS0838+13  & 0.684 & N35 & 09/04/07 & $ 0.8 $ &   54.7 & 0.00 \\
               TON392  & 0.654 & N36 & 17/10/07 & $ 0.7 $ &   57.2 & 0.07 \\
         Z101733-0203  & 1.343 & E78 & 23/03/07 & $ 0.9 $ &   79.1 & 0.05 \\
           PKS1015-31  & 1.346 & E78 & 25/03/07 & $ 1.0 $ &   26.2 & 0.05 \\
           PKS1018-42  & 1.280 & E78 & 25/03/07 & $ 1.0 $ &   11.4 & 0.05 \\
          PKS1102-242  & 1.666 & E78 & 24/03/07 & $ 0.8 $ &   18.4 & 0.05 \\
         Z124029-0010  & 2.030 & E78 & 23/03/07 & $ 1.1 $ &   20.3 & 0.05 \\
         Z133136-0002  & 2.710 & E78 & 25/03/07 & $ 1.1 $ &   16.1 & 0.05 \\
         Z143220-0215  & 2.476 & E78 & 24/03/07 & $ 1.0 $ &   14.0 & 0.05 \\
         Z144022-0122  & 2.244 & E78 & 25/03/07 & $ 0.9 $ &   49.8 & 0.05 \\
          PKSJ1511-10  & 1.513 & E78 & 23/03/07 & $ 1.3 $ &   19.7 & 0.05 \\
           PKS1524-13  & 1.687 & E78 & 24/03/07 & $ 0.7 $ &   10.3 & 0.05 \\
          HS1623+7313  & 0.621 & N35 & 09/04/07 & $ 1.5 $ &    5.7 & 0.00 \\
                3C422  & 0.942 & N36 & 18/10/07 & $ 1.0 $ &   72.3 & 0.02 \\
           MC2112+172  & 0.878 & N36 & 19/10/07 & $ 1.0 $ &   49.5 & 0.03 \\
           Q2125-4432  & 2.503 & E79 & 09/09/07 & $ 1.3 $ &   14.4 & 0.08 \\
         Z215539-3026  & 2.593 & E79 & 11/09/07 & $ 1.9 $ &   11.3 & 0.05 \\
           PKS2204-20  & 1.923 & E79 & 10/09/07 & $ 1.6 $ &   24.6 & 0.06 \\
              4C02.54  & 0.976 & N36 & 18/10/07 & $ 0.9 $ &   16.0 & 0.02 \\
         Z221139-3132  & 2.391 & E79 & 10/09/07 & $ 1.2 $ &   11.6 & 0.06 \\
         Z222702-3205  & 2.177 & E79 & 08/09/07 & $ 2.3 $ &   19.6 & 0.00 \\
           Q2225-403A  & 2.410 & E79 & 09/09/07 & $ 1.3 $ &   76.0 & 0.07 \\
           Q2225-403B  & 0.932 & E79 & 09/09/07 & $ 1.3 $ &   20.2 & 0.07 \\
           PKS2227-08  & 1.562 & E77 & 30/09/05 & $ 2.2 $ &   32.8 & 0.00 \\
         Z223048-2954  & 2.652 & E79 & 10/09/07 & $ 1.0 $ &   18.5 & 0.06 \\
         Z225950-3206  & 2.225 & E79 & 08/09/07 & $ 2.4 $ &   34.1 & 0.08 \\
         Z231751-3147  & 2.628 & E79 & 12/09/07 & $ 1.8 $ &   27.5 & 0.14 \\
         Z232755-3154  & 2.737 & E79 & 09/09/07 & $ 1.3 $ &   13.7 & 0.07 \\
         Z233451-2929  & 2.669 & E79 & 11/09/07 & $ 1.3 $ &   54.0 & 0.09 \\
          PKS2345-167  & 0.576 & N36 & 19/10/07 & $ 1.2 $ &   12.2 & 0.03 \\
           Q2348-4012  & 1.500 & E79 & 11/09/07 & $ 1.3 $ &   42.6 & 0.01 \\
          ZC2351+010B  & 0.810 & N36 & 19/10/07 & $ 1.3 $ &    7.9 & 0.03 \\
\end{longtable}

\newpage
\begin{center}
\begin{longtable}{@{}ccccccccc@{}}
  \caption[]{Continuum and line fit parameters. (1) quasar name; (2) catalogue redshift; (3) spectral
  index of the fitted power-law defined so that log $F_\lambda$ = $\alpha$ log $\lambda$ + {\it const};
  (4) quasar monochromatic flux as estimated from the fitted power-law at $\lambda=5100, 3000, 1350$ \AA{}
  (for \Hb, \Mgii{} and \Civ{} respectively); (5) monochromatic luminosity as computed from column (4) and
  assuming $H_0=70$ km/s/Mpc, $\Omega_{\rm m}=0.3$, $\Omega_{\Lambda}=0.7$; (6) fitted line; (7)
  FWHM, estimated with GH fit after modeling \FeII{} emission; (8) Black hole masses, assuming
  $f$(\Hb,\Mgii)=$1.6$ and $f$(\Civ)=$2.4$. (9) Eddington ratio, assuming $L_{\rm Bol}$/\lLl{}=$9.26$,
  $5.15$ and $3.81$ for \Hb{}, \Mgii{} and \Civ{} respectively.}\label{tab_fits}\\
 \hline \\[-2ex]
   Name              &  $z$  &$\alpha$&  log $F_\lambda$ & log $\lambda L_\lambda$ & Line & FWHM & log \Mbh{} & log \Edd{} \\
                     &       &        & [erg/s/cm$^2$/\AA] & [erg/s] &     & [km/s] & [\Msun{}] &  \\
   (1)               &   (2) & (3)    &  (4)   & (5)    & (6)     & (7)    &  (8)   & (9)      \\
 \hline
\endfirsthead
 \hline \\[-2ex]
   Name              &  $z$  &$\alpha$&  log $F_\lambda$ & log $\lambda L_\lambda$ & Line & FWHM & log \Mbh{} & log \Edd{} \\
                     &       &        & [erg/s/cm$^2$/\AA] & [erg/s] &     & [km/s] & [\Msun{}] &  \\
   (1)               &   (2) & (3)    &  (4)   & (5)    & (6)     & (7)    &  (8)   & (9)      \\
 \hline
\endhead
\\[-1.8ex] \hline
\multicolumn{9}{l}{\it Continue in next page} \\
\endfoot
\\[-1.8ex] \hline
\endlastfoot
         PKS0000-177 & 1.465 &  -1.05 & -15.67 &  45.94 & \Mgii &  6857 &    9.71 &  -1.16 \\
          Q0040-3731 & 1.780 &  -0.92 & -14.93 &  46.55 & \Civ  &  2561 &    9.16 &  -0.13 \\
             SGP2:36 & 1.756 &  -1.52 & -16.42 &  45.03 & \Civ  &  6123 &    9.13 &  -1.62 \\
             SGP5:46 & 0.955 &  -0.74 & -16.13 &  45.02 & \Mgii &  3873 &    8.77 &  -1.14 \\
            0054+144 & 0.171 &  -1.38 & -15.84 &  43.82 & \Hb   &  8220 &    8.73 &  -2.04 \\
             SGP4:39 & 1.716 &  -1.33 & -16.16 &  45.27 & \Civ  &  1852 &    8.12 &  -0.37 \\
         PKS0100-270 & 1.597 &  -1.57 & -15.24 &  46.46 & \Mgii &  3539 &    9.37 &  -0.30 \\
       LBQS0100+0205 & 0.393 &  -1.48 & -14.69 &  45.20 & \Civ  &  3966 &    8.85 &  -1.17 \\
            0110+297 & 0.363 &  -2.36 & -15.57 &  44.81 & \Hb   &  6149 &    9.13 &  -1.45 \\
         PKS0113-283 & 2.555 &  -1.10 & -15.36 &  46.50 & \Civ  &  8014 &   10.18 &  -1.20 \\
            0119-370 & 1.320 &  -1.01 & -16.22 &  45.29 & \Mgii & 10178 &    9.74 &  -1.84 \\
            0133+207 & 0.425 &  -1.63 & -14.68 &  45.27 & \Civ  &  4478 &    9.00 &  -1.25 \\
                     &       &  -2.37 & -15.57 &  44.96 & \Hb   &  8845 &    9.55 &  -1.72 \\
                3C48 & 0.367 &  -1.80 & -14.83 &  45.55 & \Hb   &  4011 &    9.24 &  -0.82 \\
        HB890137+012 & 0.260 &  -0.90 & -14.99 &  44.49 & \Civ  &  6778 &    8.93 &  -1.96 \\
           0152-4055 & 1.650 &  -0.91 & -15.91 &  45.83 & \Mgii &  4664 &    9.32 &  -0.86 \\
         PKS0155-495 & 1.298 &  -1.71 & -15.92 &  45.56 & \Mgii &  4511 &    9.16 &  -0.99 \\
          PKS0159-11 & 0.669 &  -1.88 & -14.58 &  46.19 & \Mgii &  3703 &    9.29 &  -0.49 \\
          B0204+2916 & 0.109 &  -0.81 & -14.81 &  44.40 & \Hb   &  8964 &    9.18 &  -1.91 \\
            0244+194 & 0.176 &  -2.78 & -14.77 &  44.90 & \Hb   &  4675 &    8.94 &  -1.17 \\
       KUV03086-0447 & 0.755 &  -0.99 & -14.84 &  46.06 & \Mgii &  2771 &    8.97 &  -0.30 \\
            MZZ01558 & 1.829 &  -0.66 & -16.19 &  45.31 & \Civ  &  8012 &    9.53 &  -1.74 \\
              US3828 & 0.515 &  -4.07 & -14.82 &  45.68 & \Mgii &  4066 &    9.13 &  -0.84 \\
          Q0335-3546 & 1.841 &  -1.43 & -15.76 &  45.75 & \Civ  &  3452 &    9.01 &  -0.78 \\
         PKS0348-120 & 1.520 &  -1.80 & -15.72 &  45.93 & \Mgii &  5044 &    9.43 &  -0.89 \\
          PKS0349-14 & 0.614 &  -1.50 & -14.36 &  46.32 & \Mgii &  8442 &   10.07 &  -1.14 \\
         PKS0402-362 & 1.417 &  -0.83 & -15.17 &  46.41 & \Mgii &  4251 &    9.50 &  -0.48 \\
         PKS0403-132 & 0.571 &  -2.53 & -14.96 &  45.65 & \Mgii &  3709 &    9.04 &  -0.78 \\
         PKS0405-123 & 0.574 &  -1.72 & -13.83 &  46.79 & \Mgii &  6572 &   10.07 &  -0.67 \\
          PKS0414-06 & 0.773 &   1.24 & -14.75 &  46.18 & \Mgii & 12790 &   10.36 &  -1.57 \\
         PKS0420-014 & 0.915 &  -0.36 & -14.97 &  46.13 & \Mgii &  3601 &    9.24 &  -0.50 \\
          PKS0440-00 & 0.607 &  -0.83 & -15.17 &  45.50 & \Mgii &  5129 &    9.25 &  -1.14 \\
           0624+6907 & 0.370 &  -1.08 & -13.45 &  46.38 & \Civ  &  3583 &    9.42 &  -0.56 \\
                     &       &  -3.04 & -14.28 &  46.12 & \Hb   &  3631 &    9.53 &  -0.54 \\
          PKS0710+11 & 0.768 &  -2.92 & -14.74 &  46.18 & \Mgii &  8339 &    9.99 &  -1.20 \\
       MS0824.2+0327 & 1.431 &  -1.57 & -15.82 &  45.76 & \Mgii &  2748 &    8.81 &  -0.44 \\
        MS08287+6614 & 0.610 &  -0.74 & -15.19 &  45.49 & \Mgii &  5773 &    9.35 &  -1.25 \\
          PKS0838+13 & 0.684 &  -0.37 & -16.02 &  44.78 & \Mgii &  3840 &    8.66 &  -1.27 \\
              US1867 & 0.513 &  -1.03 & -14.38 &  45.77 & \Civ  &  4222 &    9.22 &  -0.97 \\
                     &       &  -2.08 & -15.29 &  45.44 & \Hb   &  2279 &    8.69 &  -0.38 \\
            0903+169 & 0.411 &  -1.29 & -14.97 &  44.96 & \Civ  &  4989 &    8.92 &  -1.48 \\
              TON392 & 0.654 &  -1.23 & -14.46 &  46.29 & \Mgii &  4905 &    9.58 &  -0.68 \\
        MS09441+1333 & 0.131 &  -1.84 & -14.83 &  44.59 & \Hb   &  3230 &    8.40 &  -0.94 \\
            0953+415 & 0.234 &  -3.07 & -14.32 &  45.63 & \Hb   &  3606 &    9.19 &  -0.69 \\
            1001+291 & 0.330 &  -2.38 & -15.11 &  45.18 & \Hb   &  1977 &    8.33 &  -0.28 \\
                     &       &  -1.21 & -13.85 &  45.86 & \Civ  &  3072 &    9.00 &  -0.66 \\
            1004+130 & 0.240 &  -1.99 & -14.67 &  45.32 & \Hb   &  6010 &    9.45 &  -1.26 \\
        Z101733-0203 & 1.343 &  -1.12 & -16.90 &  44.61 & \Mgii &  3219 &    8.41 &  -1.19 \\
          PKS1015-31 & 1.346 &   1.50 & -16.02 &  45.50 & \Mgii &  5112 &    9.24 &  -1.13 \\
          PKS1018-42 & 1.280 &  -0.64 & -15.38 &  46.09 & \Mgii &  4353 &    9.37 &  -0.67 \\
            1058+110 & 0.423 &  -2.03 & -15.88 &  44.65 & \Hb   &  7460 &    9.20 &  -1.68 \\
            1100+772 & 0.315 &  -2.46 & -14.38 &  45.86 & \Hb   &  6980 &    9.94 &  -1.21 \\
            1116+215 & 0.177 &  -2.39 & -14.05 &  45.62 & \Hb   &  2671 &    8.90 &  -0.41 \\
            1150+497 & 0.334 &  -1.49 & -14.41 &  45.30 & \Civ  &  3966 &    8.91 &  -1.13 \\
                     &       &  -1.86 & -15.61 &  44.69 & \Hb   &  3868 &    8.65 &  -1.09 \\
            1202+281 & 0.165 &  -1.00 & -14.54 &  44.49 & \Civ  &  2560 &    8.09 &  -1.12 \\
                     &       &  -2.38 & -14.97 &  44.64 & \Hb   &  5186 &    8.86 &  -1.35 \\
            1208+322 & 0.388 &  -2.06 & -15.23 &  45.20 & \Hb   &  5279 &    9.26 &  -1.19 \\
            1216+069 & 0.331 &  -1.33 & -13.86 &  45.86 & \Civ  &  2560 &    8.84 &  -0.50 \\
             MRK0205 & 0.071 &  -1.32 & -13.56 &  44.76 & \Civ  &  2944 &    8.36 &  -1.12 \\
                     &       &  -1.19 & -14.29 &  44.61 & \Hb   &  3277 &    8.45 &  -0.97 \\
            1222+125 & 0.415 &  -2.05 & -15.56 &  44.94 & \Hb   &  7529 &    9.40 &  -1.59 \\
               3C273 & 0.158 &  -1.25 & -12.89 &  46.09 & \Civ  &  3199 &    9.16 &  -0.59 \\
            1230+097 & 0.415 &   1.38 & -14.21 &  45.73 & \Civ  &  3327 &    8.99 &  -0.78 \\
                     &       &  -2.17 & -15.21 &  45.30 & \Hb   &  4765 &    9.24 &  -1.07 \\
        Z124029-0010 & 2.030 &  -0.87 & -15.74 &  45.87 & \Civ  &  6529 &    9.65 &  -1.30 \\
          PG1302-102 & 0.286 &  -0.96 & -13.35 &  46.21 & \Civ  &  3966 &    9.41 &  -0.72 \\
            1307+085 & 0.155 &  -1.71 & -14.20 &  45.35 & \Hb   &  3494 &    9.00 &  -0.78 \\
                     &       &  -0.95 & -13.90 &  45.07 & \Civ  &  3327 &    8.63 &  -1.08 \\
            1309+355 & 0.184 &  -1.12 & -13.84 &  45.30 & \Civ  &  3199 &    8.73 &  -0.95 \\
                     &       &  -2.21 & -15.07 &  44.64 & \Hb   &  4434 &    8.72 &  -1.21 \\
        Z133136-0002 & 2.710 &  -0.68 & -16.08 &  45.84 & \Civ  &  6027 &    9.57 &  -1.25 \\
            1402+436 & 0.320 &  -1.14 & -15.74 &  44.52 & \Hb   &  2763 &    8.25 &  -0.86 \\
          PG1416-129 & 0.129 &  -1.78 & -14.18 &  44.60 & \Civ  &  3838 &    8.50 &  -1.42 \\
            1425+267 & 0.366 &  -1.24 & -14.27 &  45.54 & \Civ  &  8057 &    9.66 &  -1.64 \\
                     &       &  -2.80 & -15.17 &  45.21 & \Hb   &  8090 &    9.64 &  -1.56 \\
        Z143220-0215 & 2.476 &  -1.41 & -16.15 &  45.67 & \Civ  &  5729 &    9.43 &  -1.28 \\
        Z144022-0122 & 2.244 &  -1.01 & -15.95 &  45.77 & \Civ  &  6225 &    9.55 &  -1.30 \\
            1444+407 & 0.267 &  -2.26 & -15.95 &  44.11 & \Hb   &  2694 &    7.96 &  -0.98 \\
                     &       &  -1.40 & -13.74 &  45.74 & \Civ  &  5117 &    9.38 &  -1.16 \\
         PKSJ1511-10 & 1.513 &  -1.56 & -15.71 &  45.93 & \Mgii &  4662 &    9.36 &  -0.82 \\
             1512+37 & 0.371 &  -1.86 & -14.00 &  45.82 & \Civ  &  3966 &    9.20 &  -0.90 \\
                     &       &  -1.96 & -15.30 &  45.10 & \Hb   &  8910 &    9.65 &  -1.68 \\
          PKS1524-13 & 1.687 &  -2.05 & -15.28 &  46.13 & \Civ  &  4778 &    9.52 &  -0.91 \\
        1549+203 & 0.250 &  -2.28 & -15.14 &  44.88 & \Hb   &  1880 &    8.06 &  -0.31 \\
             3C323.1 & 0.266 &  -2.05 & -15.47 &  44.60 & \Hb   &  4734 &    8.76 &  -1.29 \\
                     &       &  -1.39 & -14.09 &  45.40 & \Civ  &  4222 &    9.02 &  -1.14 \\
         HS1623+7313 & 0.621 &  -1.50 & -15.41 &  45.29 & \Mgii &  3783 &    8.88 &  -0.98 \\
            1635+119 & 0.146 &  -1.38 & -15.04 &  44.46 & \Hb   &  5704 &    8.83 &  -1.50 \\
               3C345 & 0.594 &  -0.84 & -15.73 &  45.15 & \Hb   &  3938 &    8.98 &  -0.96 \\
                     &       &  -0.78 & -14.32 &  45.98 & \Civ  &  4350 &    9.37 &  -0.91 \\
               3C351 & 0.372 &  -1.18 & -13.88 &  45.95 & \Civ  &  5117 &    9.49 &  -1.06 \\
                     &       &  -1.69 & -15.71 &  44.69 & \Hb   &  9256 &    9.41 &  -1.85 \\
            1821+643 & 0.297 &  -2.44 & -14.02 &  46.15 & \Hb   &  5250 &    9.88 &  -0.86 \\
                     &       &  -1.24 & -13.29 &  46.30 & \Civ  &  3966 &    9.46 &  -0.68 \\
               3C422 & 0.942 &   5.16 & -15.40 &  45.74 & \Mgii &  1594 &    8.34 &   0.01 \\
          MC2112+172 & 0.878 &  -1.76 & -15.94 &  45.12 & \Mgii &  2648 &    8.49 &  -0.76 \\
          Q2125-4432 & 2.503 &  -1.04 & -16.15 &  45.68 & \Civ  &  6704 &    9.57 &  -1.41 \\
          PKS2128-12 & 0.501 &  -1.03 & -13.87 &  46.26 & \Civ  &  4733 &    9.59 &  -0.85 \\
          PKS2135-14 & 0.200 &  -0.59 & -14.14 &  45.09 & \Civ  &  4861 &    8.98 &  -1.41 \\
            2141+175 & 0.211 &  -2.28 & -14.72 &  45.14 & \Hb   &  5179 &    9.19 &  -1.18 \\
                     &       &  -0.68 & -14.00 &  45.28 & \Civ  &  4222 &    8.96 &  -1.20 \\
        Z215539-3026 & 2.593 &  -1.35 & -15.97 &  45.90 & \Civ  &  5065 &    9.45 &  -1.07 \\
            2201+315 & 0.295 &  -1.60 & -13.37 &  46.22 & \Civ  &  4733 &    9.57 &  -0.87 \\
                     &       &  -2.77 & -14.58 &  45.59 & \Hb   &  3022 &    9.01 &  -0.55 \\
          PKS2204-20 & 1.923 &  -0.97 & -16.22 &  45.34 & \Civ  &  3348 &    8.76 &  -0.94 \\
        Z221139-3132 & 2.391 &  -0.93 & -16.12 &  45.66 & \Civ  &  3756 &    9.05 &  -0.91 \\
        Z222702-3205 & 2.177 &  -1.41 & -16.11 &  45.58 & \Civ  &  6573 &    9.50 &  -1.44 \\
          Q2225-403A & 2.410 &  -1.03 & -15.70 &  46.09 & \Civ  &  5372 &    9.60 &  -1.03 \\
          Q2225-403B & 0.932 &  -0.88 & -16.13 &  44.99 & \Mgii &  4571 &    8.90 &  -1.30 \\
          PKS2227-08 & 1.562 &   0.00 &   0.00 &  46.91 & \Mgii &  1755 &    8.94 &   0.58 \\
        Z223048-2954 & 2.652 &  -1.12 & -16.22 &  45.68 & \Civ  &  5450 &    9.39 &  -1.23 \\
            2247+140 & 0.235 &  -1.48 & -15.27 &  44.68 & \Hb   &  3178 &    8.48 &  -0.93 \\
        Z225950-3206 & 2.225 &  -1.24 & -15.78 &  45.92 & \Civ  &  8110 &    9.87 &  -1.47 \\
        Z231751-3147 & 2.628 &  -1.36 & -16.34 &  45.54 & \Civ  &  2364 &    8.56 &  -0.54 \\
        Z232755-3154 & 2.737 &  -0.95 & -16.34 &  45.59 & \Civ  &  3009 &    8.81 &  -0.74 \\
        Z233451-2929 & 2.669 &  -1.18 & -16.31 &  45.59 & \Civ  &  2676 &    8.70 &  -0.63 \\
         PKS2345-167 & 0.576 &  -0.23 & -15.05 &  45.57 & \Mgii &  3350 &    8.91 &  -0.73 \\
          Q2348-4012 & 1.500 &  -1.23 & -15.82 &  45.82 & \Mgii &  3223 &    8.98 &  -0.55 \\
\end{longtable}
\end{center}
\twocolumn


\begin{thebibliography}{99}
\bibitem[\protect\citeauthoryear{Adelman-McCarthy et al.}{2007}]{adelman07} Adelman-McCarthy J.K., Ag\"ueros M.A., Allam S.S., Anderson K.S.J., Anderson S.F., Annis J., Bahcall N.A., Bailer-Jones C.A.L., et al. 2007, ApJS, 172, 634
\bibitem[\protect\citeauthoryear{Bahcall et al.}{1997}]{bahcall97} Bahcall J.N., Kirhakos S., Saxe D.H. \& Schneider D.P., 1997, ApJ, 479, 642
\bibitem[\protect\citeauthoryear{Blandford \& McKee}{1982}]{blandford82} Blandford R.D., McKee C.F., 1982, ApJ, 255, 419
\bibitem[\protect\citeauthoryear{Bon et al.}{2009}]{bon09} Bon E., Popovi\'c L.\^C., Gavrilovi\'c N., La Mura G., Mediavilla E., 2009, MNRAS in press
\bibitem[\protect\citeauthoryear{Bonning, Shields \& Salviander}{2007}]{bonning07} Bonning E.W., Shields G.A., Salviander S., ApJ Letters, 666, 13
\bibitem[\protect\citeauthoryear{Boroson \& Green}{1992}]{boroson92} Boroson T.A., \& Green R.F., 1992, ApJS, 80, 109
\bibitem[\protect\citeauthoryear{Boyce et al.}{1998}]{boyce98} Boyce P.J., Disney M.J., Blades J.C., Boksenberg A., Crane P., Deharveng J.M., Macchetto F.D., Mackay C.D., Sparks W.B., 1998, MNRAS, 298, 121
\bibitem[\protect\citeauthoryear{Boyle et al.}{2000}]{boyle00} Boyle B.J., Shanks T., Croom S.M., Smith R.J., Miller L., Loaring N., Heymans C., 2000, MNRAS, 317, 1014
\bibitem[\protect\citeauthoryear{Buzzoni et al.}{1984}]{buzzoni84} Buzzoni B., Delabre B., Dekker H., Dodorico S., Enard D., Focardi P., Gustafsson B., Nees W., et al., 1984, ESO Messenger 38, 9
\bibitem[\protect\citeauthoryear{Decarli et al.}{2008}]{decarli08a} Decarli R., Labita M., Treves A., Falomo R., 2008, MNRAS, 387, 1237
\bibitem[\protect\citeauthoryear{Decarli Treves \& Falomo}{2009a}]{decarli09a} Decarli R., Treves A., Falomo R., 2009a, MNRAS Letters, 396, 31
\bibitem[\protect\citeauthoryear{Decarli et al.}{2009b}]{decarli09b} Decarli R., Falomo R., Kotilainen J., Labita M., Scarpa R., Treves A., 2009b, Bentham Open Astronomy Journal, 2
\bibitem[\protect\citeauthoryear{Decarli et al.}{2009c}]{paperII} Decarli R., Falomo R., Treves A., Labita M., Kotilainen J.K., Scarpa R., 2009, MNRAS submitted (Paper II)
\bibitem[\protect\citeauthoryear{Drinkwater}{1987}]{drinkwater87} Drinkwater M.J., 1987, `Quasar clustering on large scale', PhD Thesis, Cambridge
\bibitem[\protect\citeauthoryear{Dunlop \& Peacock}{1990}]{dunlop90} Dunlop J. S., Peacock J. A., 1990, MNRAS, 247, 19
\bibitem[\protect\citeauthoryear{Dunlop et al.}{2003}]{dunlop03} Dunlop J.S., McLure R.J., Kukula M.J., Baum S.A., O'Dea C.P., Hughes D.H., 2003, MNRAS, 340, 1095
\bibitem[\protect\citeauthoryear{Falomo et al.}{2004}]{falomo04} Falomo R., Kotilainen J.K., Pagani C., Scarpa R., Treves, A., 2004, ApJ, 604, 495
\bibitem[\protect\citeauthoryear{Falomo et al.}{2005}]{falomo05} Falomo R., Kotilainen J.K., Scarpa R., Treves A., 2005, A\&A, 434, 469
\bibitem[\protect\citeauthoryear{Falomo et al.}{2008}]{falomo08} Falomo R., Treves A., Kotilainen J.K., Scarpa R., Uslenghi M., 2008, ApJ, 673, 694
\bibitem[\protect\citeauthoryear{Fan et al.}{2004}]{fan04} Fan X., Hennawi J.F., Richards G.T., Strauss M.A., Schneider D.P., Donley J.L., Young J.E., Annis J., et al., 2004, AJ, 128, 515
\bibitem[\protect\citeauthoryear{Ferrarese}{2006}]{ferrarese06} Ferrarese L., 2006, in Series in High Energy Physics, Cosmology and Gravitation, `Joint Evolution of Black Holes and Galaxies', ed. by M. Colpi, V. Gorini, F. Haardt, U. Moschella (New York-London: Taylor \& Francis Group), 1
\bibitem[\protect\citeauthoryear{Fine et al.}{2008}]{fine08} Fine S., Croom S.M., Hopkins P.F., Hernquist L., Bland-Hawthorn J., Colless M., Hall P.B., Miller L., et al., 2008, MNRAS, 390, 1413
\bibitem[\protect\citeauthoryear{Floyd et al.}{2004}]{floyd04} Floyd D.J.E., Kukula M.J., Dunlop J.S., et al., 2004, MNRAS, 355, 196
\bibitem[\protect\citeauthoryear{Francis et al.}{1991}]{francis91} Francis P.J., Hewett P.C., Foltz C.B., Chaffee F.H., Weymann R.J., Morris S.L., 1991, ApJ, 373, 465
\bibitem[\protect\citeauthoryear{Hamilton et al.}{2002}]{hamilton02} Hamilton T.S., Casertano S., Turnshek D.A., 2002, ApJ, 576, 61
\bibitem[\protect\citeauthoryear{H\"aring \& Rix}{2004}]{haring04} H\"aring N., Rix H.-W., 2004, ApJ Letters, 604, 89
\bibitem[\protect\citeauthoryear{Hooper et al.}{1997}]{hooper97} Hooper E.J., Impey C.D. \& Foltz C.B., 1997, ApJ Letter, 480, 95
\bibitem[\protect\citeauthoryear{Hyv\"onen et al.}{2007a}]{hyvonen07a} Hyv\"onen T., Kotilainen J.K., Orndhal E., Falomo R., 2007a, A\&A, 462, 525
\bibitem[\protect\citeauthoryear{Hyv\"onen et al.}{2007b}]{hyvonen07b} Hyv\"onen T., Kotilainen J.K., Falomo, R., Orndhal, E. Pursimo T., 2007b, A\&A, 476, 723
\bibitem[\protect\citeauthoryear{Kaspi et al.}{2000}]{kaspi00} Kaspi S., Smith P.S., Netzer H., Maoz D., Jannuzi B.T., Giveon U., 2000, ApJ, 533, 631
\bibitem[\protect\citeauthoryear{Kaspi et al.}{2007}]{kaspi07} Kaspi S., Brandt W.N., Maoz D., Netzer H., Schneider D.P., Shemmer O., 2007, ApJ, 659, 997
\bibitem[\protect\citeauthoryear{Kim et al.}{2008}]{kim08a} Kim M., Ho L.C., Peng C.Y., Barth A.J., Im M., 2008, ApJ Suppl., 179., 283
\bibitem[\protect\citeauthoryear{Kirhakos et al.}{1999}]{kirhakos99} Kirhakos S., Bahcall J.N., Schneider D.P. \& Kristian J., 1999, ApJ, 520, 67
\bibitem[\protect\citeauthoryear{Kotilainen et al.}{1998}]{kotilainen98} Kotilainen J.K., Falomo R., \& Scarpa R., 1998, A\&A, 332, 503
\bibitem[\protect\citeauthoryear{Kotilainen et al.}{2000}]{kotilainen00} Kotilainen J.K., \& Falomo R., 2000, A\&A, 364, 70
\bibitem[\protect\citeauthoryear{Kotilainen et al.}{2007}]{kotilainen07} Kotilainen J.K., Falomo R., Labita M., Treves A., \& Uslenghi M., 2007, ApJ 660 1039
\bibitem[\protect\citeauthoryear{Kotilainen et al.}{2009}]{kotilainen09} Kotilainen J.K., Falomo R., Decarli R., Treves A., Uslenghi M., Scarpa R., 2008, ApJ submitted
\bibitem[\protect\citeauthoryear{Kukula et al.}{2001}]{kukula01} Kukula M.J., Dunlop J.S., McLure R.J., et al., 2001, MNRAS, 326, 1533
\bibitem[\protect\citeauthoryear{Labita et al.}{2006}]{labita06} Labita M., Treves A., Falomo R., Uslenghi M., 2006, MNRAS, 373, 551
\bibitem[\protect\citeauthoryear{Labita et al.}{2009a}]{labita09a} Labita M., Decarli R., Treves A., Falomo R., 2009a, MNRAS, 396, 1537
\bibitem[\protect\citeauthoryear{Labita et al.}{2009b}]{labita09b} Labita M., Decarli R., Treves A., Falomo R., 2009b, MNRAS accepted (arXiv:0907.2963)
\bibitem[\protect\citeauthoryear{Madau et al.}{1998}]{madau98} Madau P., Pozzetti L., Dickinson M., 1998, ApJ, 498, 106
\bibitem[\protect\citeauthoryear{Marconi \& Hunt}{2003}]{marconi03} Marconi A., Hunt L.K., 2003, ApJ Letters, 589, 21
\bibitem[\protect\citeauthoryear{Marziani et al.}{2003}]{marziani03} Marziani P., Sulentic J.W., Zamanov R., et al., 2003 ApJS, 145, 199
\bibitem[\protect\citeauthoryear{McGill et al.}{2008}]{mcgill08} McGill K.L., Woo J.H., Treu T., Malkan M.A., 2008, ApJ, 673, 703
\bibitem[\protect\citeauthoryear{McLure \& Jarvis}{2002}]{mclure02} McLure R.J. \& Jarvis M.J., 2002, MNRAS, 337, 109
\bibitem[\protect\citeauthoryear{McLure et al.}{2006}]{mclure06} McLure R.J., Jarvis M.J., Targett T.A., Dunlop J.S., Best P.N., 2006, MNRAS, 368, 1395
\bibitem[\protect\citeauthoryear{Pagani et al.}{2003}]{pagani03} Pagani C., Falomo R. \& Treves A., 2003, ApJ, 596, 830
\bibitem[\protect\citeauthoryear{Pastorini et al.}{2007}]{pastorini07} Pastorini G., Marconi A., Capetti A., Axon D.J., Alonso-Herrero A., Atkinson J., Batcheldor D., Carollo C.M., et al., 2007, A\&A, 469, 405
\bibitem[\protect\citeauthoryear{Peng et al.}{2006a}]{peng06a} Peng C.Y., Impey C.D., Ho L.C., Barton E.J., Rix H.-W., 2006a, ApJ, 640, 114
\bibitem[\protect\citeauthoryear{Peng et al.}{2006b}]{peng06b} Peng C.Y., Impey C.D., Rix H.-W., Kochanek C.S., Keeton C.R., Falco E.E., Leh\'ar J., McLeod B.A., 2006b, ApJ, 649, 616
\bibitem[\protect\citeauthoryear{Peterson \& Wandel}{2000}]{peterson00} Peterson B.M. \& Wandel A., 2000, ApJ Letters, 540, 13
\bibitem[\protect\citeauthoryear{Phillips}{1978}]{phillips78} Phillips M.M., 1978, ApJS, 38, 187
\bibitem[\protect\citeauthoryear{Richards et al.}{2006}]{richards06} Richards G.T., Lacy M., Storrie-Lombardi L.J., Hall P.B., Gallagher S.C., Hines D.C., Fan X., Papovich C., et al., 2006, ApJS, 166, 470
\bibitem[\protect\citeauthoryear{Ridgway et al.}{2001}]{ridgway01} Ridgway S., Heckman T., Calzetti D., Lehnert M., 2001, ApJ, 550, 122
\bibitem[\protect\citeauthoryear{Savage et al.}{1984}]{savage84} Savage A., Trew A.S., Chen J.-S., Weston T., 1984, MNRAS, 207, 393
\bibitem[\protect\citeauthoryear{Schlegel Finkbeiner \& Davis}{1998}]{schlegel98} Schlegel D.J., Finkbeiner D.P., Davis M., 1998, ApJ, 500, 525
\bibitem[\protect\citeauthoryear{Schmidt}{1977}]{schmidt77} Schmidt M., 1977, ApJ, 217, 358
\bibitem[\protect\citeauthoryear{Shen et al.}{2008}]{shen08} Shen Y., Greene J.E., Strauss M.A., Richards G.T., Schneider D.P., 2008, ApJ, 680, 169
\bibitem[\protect\citeauthoryear{Tsuzuki et al.}{2006}]{tsuzuki06} Tsuzuki Y., Kawara K., Yoshii Y., Oyabu S., Tanabe T., Matsuoka Y., 2006, ApJ, 650, 57
\bibitem[\protect\citeauthoryear{Van Der Marel \& Franx}{1993}]{vandermarel93} Van Der Marel R.P. \& Franx M., 1993, ApJ, 407, 525
\bibitem[\protect\citeauthoryear{Veron-Cetty \& Veron}{2006}]{vcv06} Veron-Cetty M.P., \& Veron P., 2006, A\&A, 455, 773
\bibitem[\protect\citeauthoryear{Vestergaard \& Wilkes}{2001}]{vestergaard01} Vestergaard M., Wilkes B.J., 2001, ApJ Suppl., 134, 1
\bibitem[\protect\citeauthoryear{Zhan \& Chen}{1989}]{zhan89} Zhan Y., Chen J.-S., 1989, ChA\&A, 13, 139
\end{thebibliography}
\end{document}